\documentclass[a4paper,notitlepage,twocolumn,superscriptaddress,longbibliography,nofootinbib]{revtex4-1}
\pdfoutput=1
\usepackage[utf8]{inputenc}
\usepackage{amsmath,amsthm,amssymb}
\usepackage{scalerel}
\usepackage{mathtools}
\usepackage{graphicx}
\usepackage{subfigure}
\usepackage[colorlinks = true,
            linkcolor = blue,
            urlcolor  = blue,
            citecolor = blue,
            anchorcolor = blue]{hyperref}
\usepackage{mathrsfs}
\usepackage{bbold}
\usepackage[]{units}
\usepackage{bm}
\usepackage{braket}
\usepackage{color}
\usepackage{makecell}
\usepackage{enumitem}
\usepackage{upgreek}
\usepackage{blindtext}
\usepackage{graphics}
\usepackage{verbatim} 

\usepackage[dvipsnames]{xcolor}
\usepackage{bbm}
\usepackage[bb=boondox]{mathalfa}
\usepackage{array}
\usepackage{multirow}
\usepackage{tabularx}
\usepackage{float}
\usepackage{graphicx}
\usepackage{dcolumn}
\usepackage{bm}
\usepackage{amsmath,amsfonts,amssymb}
\usepackage{slashed}
\usepackage{braket,xcolor}
\usepackage{verbatim}
\usepackage{multirow}
\usepackage{amsfonts,mathtools,resizegather}
\usepackage[export]{adjustbox}

\newcommand{\kee}[1]{\vert#1)}
\newcommand{\bee}[1]{(#1\vert}
\newcommand{\ipr}[2]{(#1 \vert #2)}

\newcommand{\tr}[1]{\text{Tr}(#1)}

\newcommand{\sg}[1]{\text{sgn}(#1)}

\DeclareMathOperator{\sech}{sech}

\begin{document}

\title{A Lanczos approach to the Adiabatic Gauge Potential}
\author{Budhaditya Bhattacharjee}
\affiliation{Centre for High Energy Physics, Indian Institute of Science, C.V. Raman Avenue, Bangalore-560012, India}

\begin{abstract}
The Adiabatic Gauge Potential (AGP) is the generator of adiabatic deformations between quantum eigenstates. There are many ways to construct the AGP operator and evaluate the AGP norm. Recently, it was proposed that a Gram-Schmidt-type algorithm can be used to explicitly evaluate the expression of the AGP \cite{hatomura2021controlling}. We employ a version of this approach by using the Lanczos algorithm to evaluate the AGP operator in terms of Krylov vectors and the AGP norm in terms of the Lanczos coefficients. It has the advantage of minimizing redundancies in evaluating the nested commutators in the analytic expression for the AGP operator. The algorithm is used to explicitly construct the AGP operator for some simple systems. We derive an integral transform relation between the AGP norm and the autocorrelation function of the deformation operator. We present a modification of the Variational approach to derive the regulated AGP norm. Using this, we approximate the AGP to varying degrees of success. Finally, we compare and contrast the quantum-chaos-probing capacities of the AGP and K-complexity in view of the Operator Growth Hypothesis.
\end{abstract}
\maketitle
\section{Introduction} \label{intro}
The study of quantum chaos has been the focus of a large body of research over the last century. These works have led to various insights on thermalization, information scrambling, and many other phenomena in many-body physics as well as holography. In classical physics, chaos is a well-understood phenomenon. It is usually understood in terms of classical phase space trajectories. When an infinitesimal initial perturbation in the phase space variables ends up causing an exponentially large change at late times, it is understood as (classical) chaotic dynamics. This is the well-known ``butterfly effect'' \cite{Roberts:2014ifa,Roberts:2016wdl}. 

This definition does not carry over to quantum mechanics since trajectories are no longer well-defined objects. In quantum mechanics, there is no first-principle \textit{definition} of chaos yet. Instead, there are various probes of chaotic quantum dynamics. Some of the probes focus on the nature of the eigenvalues (such as spectral statistics \cite{PhysRevLett.52.1, Berry_Tabor}, eigenstate thermalization hypothesis \cite{Srednicki:1994mfb, Deutsch_2018, DAlessio:2015qtq}, etc.). Other probes focus on the spreading of operators within the system. These include probes such as operator size \cite{Roberts:2014isa,Roberts:2018mnp,Qi:2018bje, Lensky:2020ubw, Schuster:2021uvg}, out-of-time-ordered-correlators (OTOCs) \cite{Rozenbaum:2016mmv,Hashimoto:2017oit,Nahum:2017yvy, Xu:2019lhc, Zhou:2021syv, Gu:2021xaj} and Krylov complexity \cite{PhysRevX.9.041017, Barbon:2019wsy, Rabinovici:2020ryf, Bhattacharjee:2022vlt}. Part of our focus in this article will be on (the machinery of) Krylov complexity, which has found wide-ranging applications from a few body systems to quantum field theories to open quantum systems. \cite{Barbon:2019wsy, Bhattacharjee:2022vlt, Avdoshkin:2019trj, Dymarsky:2019elm, Jian:2020qpp, Rabinovici:2020ryf, Cao:2020zls, Dymarsky:2021bjq, Kar:2021nbm, Caputa:2021sib, Kim:2021okd, Magan:2020iac, Caputa:2021ori,  Bhattacharya:2022gbz, Rabinovici:2022beu, Muck:2022xfc, Liu:2022god, Patramanis:2021lkx, Caputa:2022eye, Rabinovici:2021qqt, Fan:2022xaa, Fan:2022mdw, Trigueros:2021rwj,  Hornedal:2022pkc, Balasubramanian:2022tpr, Bhattacharjee:2022qjw, Heveling:2022hth, Caputa:2022yju, Balasubramanian:2022dnj, Afrasiar:2022efk, Guo:2022hui, Bhattacharjee:2022lzy}. 

Another probe that has been considered in the study of quantum chaos is the Adiabatic Gauge Potential \cite{PhysRevX.10.041017} (AGP). The Adiabatic Gauge Potential (AGP) is the generator of adiabatic deformations between quantum eigenstates. In a precise sense, the distance between two eigenstates (i.e., the Fubini-Study metric \cite{KOLODRUBETZ20171, PhysRevA.36.3479}) is the Frobenius norm of the AGP (operator) \cite{KOLODRUBETZ20171, Berry_2009, Demirplak_2005, Demirplak_2003, selspol}. It has been observed that the (regularised) AGP demonstrates different scaling behaviors (with system size) for chaotic and integrable systems. This norm scales exponentially with the system size for chaotic systems following ETH \cite{KOLODRUBETZ20171}. It has the potential to serve as a probe of quantum chaos. It has also been considered in the study of thermalization \cite{Nandy:2022hcm}. The Adiabatic Gauge Potential itself has a rich history of being studied in various ways \cite{PhysRevLett.74.1732,PhysRevX.4.021013, PhysRevA.88.040101, PhysRevE.99.050102, PhysRevB.99.224202, MpandeyBU}. It is a unique probe since it straddles the fine line between operator dynamics and eigenstate statistics, incorporating both the information about eigenstates as well as the specific operator choice. 

We study the AGP by utilizing the Lanczos algorithm \cite{viswanath1994recursion}, which is a central ingredient in the formalism of Krylov complexity \cite{PhysRevX.9.041017}. The language of the Krylov basis (generated from the Lanczos algorithm) provides a way to construct the AGP operator by evaluating the minimum number of nested commutators (of the Hamiltonian with the deforming operator). A similar prescription was proposed in \cite{hatomura2021controlling}. The prescription described there is somewhat general. In our article, we focus on the Lanczos algorithm (which is an instance of the general algorithm in \cite{hatomura2021controlling}) and utilize it to derive an explicit expression for the AGP operators of various systems and deformations. We write the AGP operator in the Krylov language. We also study the equivalence between the variational approach towards constructing the AGP \cite{selspol} and the Krylov construction. We generalize the variational approach to regularised AGP and demonstrate a method (based on this approach) to evaluate the AGP norm. We discuss how well this approach approximates the AGP for chaotic and integrable systems\footnote{When talking about chaotic and integrable systems, we shall resort to the statement of the Operator Growth Hypothesis \cite{PhysRevX.9.041017}. Systems demonstrating linear growth of Lanczos coefficients will be considered as chaotic and anything else will be considered as integrable or non-chaotic (We shall use these terms interchangeably). There are exceptions to this, which we shall ignore for the purpose of this manuscript. For an example of such exceptions, see \cite{Bhattacharjee:2022vlt}}. Finally, we compare and contrast AGP and K-complexity as probes of chaos and phase transitions. 

\section{Adiabatic Eigenstate Deformation}
Consider a Hamiltonian $H(\lambda)$, which is a function of a tunable parameter $\lambda$. Instantaneously in $\lambda$, the eigenstates of H are defined as follows
\begin{align}
	H(\lambda)\ket{n(\lambda)} = E_{n}(\lambda)\ket{n(\lambda)}\,.
\end{align}
where $\ket{n(\lambda)}$ is the eigenstate with corresponding eigenvalue $E_{n}(\lambda)$. The adiabatic gauge potential, or AGP, is introduced by transforming to the moving frame of $\lambda$, characterized by the instantaneous energy basis. In this moving frame, the Hamiltonian is diagonal. This can be observed by considering the time evolution of a state $\ket{\psi}$ in the moving frame where it becomes $\ket{\tilde{\psi}} = U^\dagger (\lambda(t))\ket{\psi}$. Here $U$ is the unitary operator that transforms between the bases. For the sake of generality, the parameter $\lambda$ is taken to be a function of time $t$. 

In the moving frame, this state evolves as 
\begin{align}
	i \hbar \partial_{t}\ket{\tilde{\psi}} = \left(U^\dagger H U - i \hbar\, \dot{\lambda}\, U^\dagger \partial_{\lambda} U \right)\ket{\tilde{\psi}}\,.
\end{align}
The first term, $U^\dagger H U$, is diagonal by definition. The off-diagonal contribution comes from the second term $ i \hbar\, \dot{\lambda}\, U^\dagger \partial_{\lambda} U$. This term is responsible for excitations between eigenstates, quantified by the velocity $\dot{\lambda}$. The AGP in the moving frame is defined as this off-diagonal term
\begin{align}
	\tilde{A}_{\lambda} = i \hbar\, U^\dagger \partial_{\lambda} U\,.
\end{align}
It is clear that if this extra term is added to the Hamiltonian $H$, then the shifted Hamiltonian $H'$ will naturally become diagonal in the moving frame. Thus the system can demonstrate transition-less driving at arbitrary rates. Formally, we denote this as 
\begin{align}
	H' = H + \dot{\lambda}A_\lambda
\end{align}
where we have denoted the AGP in the lab frame as $A_\lambda = U \tilde{A}_\lambda U^\dagger = i \partial_{\lambda}$.

The object we seek to study is the AGP in the lab frame. The matrix elements of the same, in the eigenbasis of the original Hamiltonian $H(\lambda)$, can be written via the Feynmann-Hellmann theorem \cite{PhysRev.56.340}
\begin{align}
	\bra{m}A_\lambda \ket{n} =  -i\frac{\bra{m}\partial_\lambda H \ket{n}}{E_{m} - E_{n}}\;\;\; \forall \;\; m \neq n\,.\label{AGPnorm5}
\end{align}
An equivalent expression for the AGP can be obtained from the commutator equation $[H, i \partial_\lambda H + [H, A_\lambda]] = 0$ \cite{KOLODRUBETZ, PhysRevA.88.040101}. It is worth noting that the AGP has inherent gauge freedom associated with the choice of the phase of the eigenstates of $H$. In the matrix form, this translates into the freedom to choose the diagonal elements of the AGP. In the operator picture, the gauge freedom is realized by adding any operator $K$ to the AGP such that $[H, K] = 0$ (i.e., shifting by any conserved charge in the theory). 

The regularized AGP \cite{PhysRevX.10.041017} is defined in terms of the operator $A_\lambda$ by taking the expectation value of the same between the eigenstates of the Hamiltonian $H(\lambda)$ and taking the square of the magnitude of the same. Finally, the eigenstates are summed over. In other words, it is written as
\begin{align}
	\vert \vert A_{\lambda} \vert \vert ^{2} &= \frac{1}{\mathcal{D}}\sum_{m \neq n}\vert\bra{m}A_{\lambda}\ket{n}\vert^{2} \notag \\ &= 
	\frac{1}{\mathcal{D}}\sum_{m \neq n}\frac{\omega^{2}_{m n}}{(\mu^{2} + \omega^{2}_{m n})^{2} }\vert \bra{m}\partial_{\lambda} H \ket{n}\vert^{2}\,. \label{agporg2}
\end{align}
where $\mathcal{D}$ is the dimension of the Hilbert space and $\omega_{m n} = E_{m} - E_{n}$, i.e. the difference between eigenvalues.

\section{Lanczos Algorithm and Krylov complexity}
The notion of Krylov complexity involves choosing an appropriate basis (in the Hilbert space of operators or states) to describe the time evolution of an operator. While there can exist infinite possible such bases, there is a unique choice that corresponds to the minimal basis with respect to some appropriately defined cost functional. We only discuss operator complexity in this paper. For state complexity and its applications, the reader is directed to \cite{PhysRevX.9.041017, Balasubramanian:2022tpr, Bhattacharjee:2022qjw, Afrasiar:2022efk}.

Arriving at the appropriate basis (known as Krylov basis) involves writing the time-evolved operator $\mathcal{O}(t)$ in terms of nested commutators. This follows from the Baker-Campbell-Hausdorff formula
\begin{align}
	\mathcal{O}(t) &= e^{i H t/ \hbar}\mathcal{O} e^{-i H t/ \hbar} \notag \\ &= \mathcal{O} + \frac{i t}{\hbar} [H,\mathcal{O}] + \frac{(i t)^{2}}{2 !\,\hbar^{2}}[H, [H, \mathcal{O}]]  \notag \\ &+ \frac{(i t)^{3}}{3 ! \, \hbar^{3}}[H, [H, [H, \mathcal{O}]]] + \frac{(i t)^{4}}{4 ! \, \hbar^{4}}[H,[H, [H, [H, \mathcal{O}]]]] \notag \\ &+ \cdots~~.
\end{align}
Each term in the BCH expansion is then chosen as the basis elements of our initial basis\footnote{This is not the Krylov basis, in general.} $\mathfrak{D}$. The elements of the basis are then written as
\begin{align}
	\mathfrak{D} = \{\mathcal{O}, [H,\mathcal{O}], [H, [H, \mathcal{O}]],[H, [H, [H, \mathcal{O}]]], \cdots  \}
\end{align}
This basis is conveniently represented in the language of the Louivillian super-operator $\mathcal{L}$, which is defined as
\begin{align}
	\mathcal{L}(\ast) = [H, \ast]
\end{align}
which allows us to write the basis $\mathfrak{D}$ as
\begin{align}
	\mathfrak{D} = \text{span}\{\mathcal{L}^{n}\mathcal{O}\}_{n = 0}^{+\infty}\label{spanD}
\end{align}
The algorithm for generating the Krylov basis from $\mathfrak{D}$ is known as the Lanczos algorithm \cite{viswanath1994recursion}. It is a Gram-Schmidt-like recursive orthonormalization protocol. We start with the normalized initial operator $\mathcal{O}_0$. Normalization is defined via the Wightman inner product (at inverse temperature $\beta = 0$), written as $(\mathcal{O}| \mathcal{O}') = \frac{1}{\mathcal{N}}\mathrm{Tr}(\mathcal{O}^{\dagger}\mathcal{O}' )$, where $\mathcal{N} = \tr{\mathbf{1}}$ is the dimension of the Hilbert space of operators. 

The algorithm proceeds as follows.
\begin{itemize}
	\item Start with the definition $A_{0} \equiv \mathcal{O}_{0}$, which we assume to be normalized $(\mathcal{O}_{0}| \mathcal{O}_{0}) = 1 $.  This is the $0^\text{th}$ Krylov vector.
	\item  Define $A_{1} = \mathcal{L}\mathcal{O}_{0}$, and normalize it with $b_{1} = \sqrt{(A_{1}|A_{1})}$. Define the normalized operator $\mathcal{O}_{1} = b_{1}^{-1}A_{1}$. This is the $1^\text{st}$ Krylov vector.
	\item From this, given $\mathcal{O}_{n-1}$ and $\mathcal{O}_{n-2}$, we can construct the following operators
	\begin{equation}
		A_{n} = \mathcal{L}\mathcal{O}_{n-1} - b_{n-1}\mathcal{O}_{n-2}\,.\label{recAN}
	\end{equation}
	This can be normalized as $b_{n} = \sqrt{(A_{n}|A_{n})}$ and the $n^{\mathrm{th}}$ Krylov vector is given by $\mathcal{O}_{n} = b_{n}^{-1}A_{n}$. 
	\item Terminate the algorithm when $b_n$ hits zero.
\end{itemize}

The features of this Krylov space formed by $\text{span}\{\mathcal{O}_n\}$ are extensive and cannot fit into this tiny section. We refer the readers to \cite{Rabinovici:2021qqt, Rabinovici:2020ryf, Rabinovici:2022beu, Barbon:2019wsy} for an exhaustive review. For our purposes, it will suffice to know that the dimensions of the Krylov subspace thus formed is given by $K \leq \mathcal{D}^2 - \mathcal{D} + 1$, where $\mathcal{D}$ is the dimension of the Hilbert space generated by the Hamiltonian $H$. This also implies that the span in \eqref{spanD} is up to $n = K$ in reality.   

The time evolved operator $\mathcal{O}(t)$ is then written in terms of the Krylov basis operators $\mathcal{O}_n$ as
\begin{align}
	\mathcal{O}(t) = \sum_{n = 0}^{K}i^{n}\psi_{n}(t) \mathcal{O}_n\,. \label{Ot1}
\end{align}
The time-dependent coefficients thus obtained $\psi_n (t)$ are known as the Krylov basis wavefunctions and contain the time evolution information. These wavefunctions satisfy the following recursive ``Schr\"odinger equation'', which can be seen from the Lanczos algorithm
\begin{align}
	\partial_t \psi_{n}(t) = - b_{n+1}\psi_{n+1}(t) + b_{n}\psi_{n-1}(t)\,. \label{sch}
\end{align}
where $\psi_n (0) = \delta_{n 0}$  and $b_0 = \psi_{-1} = 0$. 

We focus only on Hermitian initial operators $\mathcal{O}^\dagger_0 = \mathcal{O}_0$. The mechanism of the Krylov construction is ill-understood for non-Hermitian operators. A few facts for Hermitian $\mathcal{O}_0$ follow from the Lanczos algorithm. Firstly, one can note that $\mathcal{O}_{ 2 n}$ are Hermitian for all $n$ and $\mathcal{O}_{2n + 1}$ are anti-Hermitian. This is the motivation behind the $i^{n}$ in the expression \eqref{Ot1}, which ensures that $\psi_{n}(t)$ are real functions. 

Secondly, one can note that for Hermitian initial operator $\mathcal{O}_0$, the Louivillian super-operator can be realized in the following matrix form
\begin{align}
	L_{m n} = \bee{\mathcal{O}_m}\mathcal{L}\kee{\mathcal{O}_n} = 
	\begin{pmatrix}
		0 & b_{1} & 0 & 0 & \cdots\\
		b_{1} & 0 & b_{2} & 0 & \cdots \\
		0 & b_{2} & 0 & b_{3} & \cdots \\
		0 & 0 & b_{3} & 0 & \cdots \\
		\vdots & \vdots & \vdots & \vdots & \ddots 
	\end{pmatrix}
\end{align}
where the matrix element $\bee{\mathcal{O}_m}\mathcal{L}\kee{\mathcal{O}_n}$ is given by the inner product $\frac{1}{\mathcal{N}}\tr{\mathcal{O}^\dagger_m \mathcal{L} \mathcal{O}_n}$.

The machinery of Krylov complexity with the Lanczos coefficients $b_n$, the wavefunctions $\psi_{n}(t)$ and the basis operators $\mathcal{O}_{n}$ together provide a complete set of tools to describe the nature of time evolution of the Hermitian operator $\mathcal{O}(t)$. We defer the description of the actual quantity known as Krylov complexity to the next section.

\subsection{Universal Operator Growth Hypothesis}
In this section, we will discuss a proposal for using the technology described above to probe chaotic dynamics. The proposal \cite{PhysRevX.9.041017} is known as the ``Universal Operator Growth Hypothesis''. The fundamental claim is that for chaotic systems, the Lanczos coefficients $b_n$ show asymptotically linear growth with $n$. For other systems, the growth is sublinear. In other words, the maximum possible growth of the Lanczos coefficients (under some assumptions) is linear in $n$. The Krylov complexity is defined as
\begin{align}
	K(t) = \bee{\mathcal{O}(t)}n\kee{\mathcal{O}(t)} = \sum_{n = 0}^{K}n \vert \phi_{n}(t) \vert^{2}\,. \label{Ot}
\end{align}
shown an exponential growth $K(t) \sim e^{2 \alpha t}$ for chaotic systems. The coefficient $\alpha$ is system dependent and corresponds to the slope of linear growth of $b_n$ (i.e., $b_n \sim \alpha n$). It also serves as an upper bound to the Lyapunov constant $\lambda \leq 2\alpha$. 

Intricately related to K-complexity, there are a few other quantities that embody the operator growth hypothesis equally well. These quantities contain exactly the same amount of information as K-complexity and the Lanczos coefficients do. These are the autocorrelation function. 
\begin{align}
	\mathcal{C}(t) = \ipr{\mathcal{O}(t)}{\mathcal{O}} = \frac{\tr{\mathcal{O}^\dagger (t) \mathcal{O}}}{\tr{\mathbf{1}}} = \bee{\mathcal{O}}e^{i \mathcal L t}\kee{\mathcal{O}}\,.\label{AutoC}
\end{align}
The autocorrelation function has the following Taylor series expansion in $t$
\begin{align}
	\mathcal{C}(-i t) = \sum_{n} \frac{m_{2 n}}{(2 n)!}t^{2 n}
\end{align}
An equivalent statement of the Operator Growth Hypothesis for chaotic systems is that the moments $m_{2 n}$ go as $n^{2 n } e^{O(n)}$ asymptotically. The Lanczos coefficients can be obtained from the moments via a recursive algorithm \cite{viswanath1994recursion}.

Certain transforms of the autocorrelation function are also of interest. These are the spectral function $\Phi(\omega)$ and the Green's function $G(z)$ defined as
\begin{align}
	\Phi(\omega) &= \int_{-\infty}^{\infty}\mathcal{C}(t) e^{i \omega t}\mathrm{d}t\,.\label{spect} \\
	G(z) &= \bee{\mathcal{O}}\frac{1}{z - \mathcal{L}}\kee{\mathcal{O}} = i \int_{0}^{\infty}e^{-i z t} \mathcal{C}(t)\mathrm{d}t\,.\label{Kgreen}
\end{align}

Finally, the exponential growth of Krylov complexity(i.e., chaotic systems) corresponds to the presence of a series of complex poles for the autocorrelation function\footnote{For systems that do not demonstrate exponential growth of $K(t)$, the autocorrelation function is analytic in the complex plane. There are a few known exceptions to this fact, though.}, with the pole closest to the real axis given by $t = \frac{\pm i \pi}{2 \alpha}$. Additionally, it also implies a large-$\omega$ fall-off for $\Phi(\omega)$, of the form $e^{-\omega\pi/2 \alpha}$. This behavior is also consistent with ETH, as we shall see.

\section{The Formalism}
In this section, we will consider the Regularized Adiabatic Gauge Potential \cite{PhysRevX.10.041017} in the Krylov/Lanczos language. The regularized AGP operator, in a particular choice of gauge (where its' diagonal elements are $0$), is given by the following expression (for the Hamiltonian $H(\lambda)$)
\begin{align}
	A_\lambda = -\frac{1}{2}\int_{-\infty}^{\infty} \sg{t}e^{-\mu \vert t \vert}\left(\partial_{\lambda}H\right) (t)\mathrm{d}t\,.\label{AGP}
\end{align}
Here $\mu$ is an infinitesimal regulator. When evaluating the AGP norm to study chaotic and integrable dynamics, it is convenient to choose $\mu = L  2^{-L}$ where $L$ is the system size. Unless otherwise mentioned, the operators and eigenstates are to be taken as functions of the adiabatic parameter $\lambda$. 

The machinery of Krylov complexity serves as a natural approach to characterizing the $\partial_\lambda H (t)$ term, provided this term is Hermitian. For the rest of this work, we shall assume that this term is normalized appropriately by the trace norm $(A|B)$ at $t = 0$. One can regain the results for the unnormalized AGP operator by multiplying the final result by its trace norm\footnote{For the AGP norm, one has to multiply the final result by $\vert \vert \partial_\lambda H \vert \vert^{2}$ to obtain the unnormalized one.} $\vert \vert \partial_\lambda H \vert \vert$.

Using \eqref{Ot1} the operator $A_\lambda$ can be written as
\begin{align}
	A_\lambda = - \frac{1}{2}\int_{-\infty}^{\infty}\sg{t}e^{-\mu \vert t \vert}\sum_{n = 0}^{K}i^{n}\psi_{n}(t) \mathcal{O}_n \mathrm{d}t\,.\label{Al}
\end{align}
The knowledge of the Krylov basis vectors and the Krylov wavefunctions is enough to construct the AGP. The Lanczos algorithm minimizes the number of terms (i.e., nested commutators) that need to be calculated (analytically) when evaluating $\mathcal{O}(t)$. Therefore, in that respect, this process achieves the evaluation of $A_\lambda$ with the minimum number of analytic evaluations.

The only contribution in \eqref{Al} is from the odd indexed Krylov vectors, i.e., only from operators of the form $\mathcal{O}_{2 k + 1}$. This is due to the fact that $\psi_{n}(-t) = (-1)^{n}\psi_{n}(t)$. Therefore \eqref{Al} reduces to 
\begin{align}
	A_\lambda = -\sum_{n = 0}^{M}\alpha_{2 n + 1}\mathcal{O}_{2 n + 1}\,.\label{AGPC1}
\end{align}
where
\begin{align}
	\alpha_{2 n + 1} = \int_{0}^{\infty}i^{2 n + 1} e^{-\mu t} \psi_{2 n + 1}(t)\mathrm{d}t\,. \label{al}
\end{align}
and $M = \frac{K-1}{2}$ if $K$ is odd or $M = \frac{K}{2}-1$ if $K$ is even. Using this, the AGP norm can be written as (detailed derivation in Appendix \ref{appA})
\begin{align}
	\vert \vert A_\lambda \vert \vert^{2} = -\sum_{n = 0}^{M}\alpha_{2 n + 1}^{2}\label{eq30}
\end{align}

This construction provides an alternate representation of the AGP operator. A nice consistency check of this expression is to check the gauge choice of $A_\lambda$. The gauge choice in this section corresponds to the diagonal elements of $A_\lambda$ being set to $0$. This is reflected via the fact that $\bra{n}\mathcal{O}_{2 n + 1}\ket{n} = 0 \;\; \forall \; n$. This follows from the Lanczos algorithm (described in the previous section) and the fact that $\bra{n}[H, Q]\ket{n} = 0$ for any operator $Q$ and eigenstate $\ket{n}$ of the hermitian Hamiltonian $H$.

\subsection{The response function}
The response function(as defined in \cite{PhysRevX.10.041017}) provides another representation of the AGP norm. It is interesting since it possesses a direct interpretation within the ETH ansatz \cite{PhysRevE.50.888}. The response function is defined as
\begin{align}
    \overline{\vert f_\lambda (\omega) \vert}^{2} &= \frac{1}{\mathcal{D}}\sum_{n}\sum_{n\neq m}\vert \bra{m}\partial_\lambda H \ket{n} \vert^{2}\delta(\omega_{m n} - \omega) \notag \\
    &= \frac{1}{\mathcal{D}}\sum_n \int_{-\infty}^{\infty}\frac{\mathrm{d}t}{4\pi}e^{i \omega t}\bra{n}\{\partial_\lambda H(t), \partial_\lambda H \}\ket{n}_c
\end{align}
where the connected component of the expectation value is defined as $\bra{n}A B\ket{n}_{c} = \bra{n}A B\ket{n} - \bra{n}A\ket{n}\bra{n}B\ket{n}$. The sum is over the instantaneous eigenstates of $H$. This expression reduces to 
\begin{align}
   \overline{\vert f_\lambda (\omega) \vert}^{2} = \frac{1}{\mathcal{D}}&\sum_{n}\int_{-\infty}^{\infty}\frac{\mathrm{d}t}{2 \pi}e^{i \omega t}\Big(\bra{n}\partial_{\lambda}H(t)\partial_\lambda H(0)\ket{n} \notag \\ &-\bra{n}\partial_\lambda H(t) \ket{n}\bra{n}\partial_\lambda H(0) \ket{n}\Big)\,.\label{rep}
\end{align}
Using \eqref{AutoC} and $\bra{n}\partial_\lambda H(t) \ket{n} = \bra{n}\partial_\lambda H(0)\ket{n}$, we obtain the following expression
\begin{align}
    \overline{\vert f_\lambda (\omega) \vert}^{2} = \int_{-\infty}^{\infty}\frac{\mathrm{d}t}{2\pi}e^{i\omega t}\mathcal{C}(t) - \frac{1}{\mathcal{D}}\sum_n \bra{n}\partial_\lambda H(0)\ket{n}^{2}\delta(\omega)\,.\label{f40}
\end{align}
The response function $\overline{\vert f_\lambda (\omega) \vert}^{2}$ is nearly the same as the spectral function defined in \eqref{spect}. We denote the extra term (proportional to $\delta(\omega)$) by $\varPhi_0$. The response function is written as
\begin{align}
    \overline{\vert f_\lambda (\omega) \vert}^{2} = \Phi(\omega) - \varPhi_{0}\delta(\omega)
\end{align}
For large $\omega$, the only contribution comes from $\Phi(\omega)$. It was demonstrated in \cite{PhysRevE.100.062134, PhysRevB.102.075127, PhysRevLett.125.070605} that the averaged over eigenstates squared of the ETH spectral function $f_{\lambda}(\omega)$ decays as $\sim e^{-\omega^n},\, n > 1$ for integrable systems and as $\sim e^{-\omega}$ for chaotic systems. This behavior is straightforward to note from the properties of the function $\Phi(\omega)$. 

It is known \cite{PhysRevX.9.041017, viswanath} that the decay rate of the spectral function $\Phi(\omega)$ is bounded above by $e^{-k |\omega|}$. The exact decay rate goes as $\sim e^{-|\omega/ \omega_0|^{1/\delta}}$ (where $\omega_0 = \frac{2}{\pi}\alpha$). The Lanczos coefficients for the same system grow (asymptotically) as $b_{n} \sim n^{\delta}$ (where $\delta \leq 1$). Therefore, for integrable systems (as identified in terms of the Operator Growth Hypothesis), it is expected that the spectral function, and by extension the response function, will decay faster than $\sim e^{-\omega}$ for large $\omega$. For chaotic systems (again, in the Operator Growth Hypothesis definition), it is expected that the spectral function (and thus the response function) will decay as $\sim e^{-\omega}$. These results follow from the statement of the Operator Growth Hypothesis, which states that chaotic systems exhibit linear growth of Lanczos coefficients (i.e., $\delta = 1$) and non-chaotic systems exhibit sublinear growth (i.e., $\delta < 1$). Therefore, we find evidence that within the ETH regime, the Operator Growth Hypothesis is compatible with ETH\footnote{In some cases \cite{Cao:2020zls}, the decay of the spectral function is known to be of the form $e^{-\omega \log\omega}$}.  

Returning to the regularized AGP norm, with a bit of algebra, one can show that
\begin{align}
    \vert \vert A_\lambda \vert \vert^{2} = \frac{1}{2}\int_{0}^{\infty} \mathrm{d}t \left(\frac{1}{\mu} - t\right)\mathcal{C}(t)e^{-\mu t}\,. \label{AGPC2}
\end{align}
The regularized AGP norm can also be expressed in terms of the Krylov Greens' function
\begin{align}
    \vert \vert A_\lambda \vert \vert^2 = -\frac{i}{2}\left(\frac{1}{\mu} + \partial_\mu \right)G(i \mu)\,.\label{AGPGreen}
\end{align}
It is also possible to describe the AGP norm solely in terms of the Lanczos coefficients. The formal expression is discussed in Appendix \ref{appA}. In the next section, we introduce a matrix equation based on the gauge constraint \cite{selspol}, which allows us to evaluate the regularized AGP norm solely in terms of the Lanczos coefficient.

\section{A Matrix Equation}
We shall now consider the gauge condition that has to be satisfied by the AGP operator \cite{selspol, KOLODRUBETZ}
\begin{align}
	[H, i \partial_\lambda H + [H,A_\lambda]] = 0\,. \label{gag}
\end{align}
This gauge condition does not hold exactly for the regularized AGP operator. When the regulator is added to the AGP operator, it may be expressed as \eqref{AGP}. It can be seen that the regularization implies that gauge constraint is modified to
\begin{align}
	[H, i\partial_\lambda H + [H,[H,A_\lambda]] + \mu^2 A_\lambda = 0\label{reggauge}
\end{align}
This expression is derived in Appendix \ref{appC}. We consider the implication of this identity vis-à-vis \eqref{AGPC1}. We denote this expression as
\begin{align}
	A_\lambda = \sum_{n = 0}^{M}\alpha_{2 n + 1}\mathcal{O}_{2 n + 1}\label{AGPC4}
\end{align}
where $\alpha_{k} = -i^{k}\int_{0}^{\infty}e^{-\mu t}\psi_{k}(t)\mathrm{d}t$. 

From the recursion relation \eqref{sch}, the relation for $\alpha_{n}$ follows
\begin{align}
	i \mu \alpha_n + b_{n}\alpha_{n - 1} + b_{n + 1}\alpha_{n + 1} + i \delta_{n, 0} = 0\,. \label{arec}
\end{align}
Using this, one can demonstrate that the gauge condition \eqref{reggauge} is satisfied. 

The expression of the AGP norm \eqref{eq30} can be equivalently written as
\begin{align}
	\vert \vert A_\lambda \vert \vert^2 = - \frac{i b_1}{2 \mu}\partial_\mu \alpha_{1}
\end{align}
This expression is simply a repackaging of \eqref{AGPGreen}.

\subsection{A set of linear equations}
The constraint \eqref{reggauge} is useful to determine the coefficients $\alpha_{2 n + 1}$. Computationally, it is somewhat straightforward to evaluate the Krylov vectors $\mathcal{O}_{2 n + 1}$ and Lanczos coefficients $b_n$. On the other hand, evaluating the wavefunctions $\psi_{2n+1}$ (and consequently $\alpha_{2 n + 1}$) becomes a very tedious process after a few wavefunctions. To circumvent this problem, one can instead solve a set of $M + 1$ linear equations to determine the $M + 1$ coefficients $\alpha_{2n + 1}$. 

Inserting the expression \eqref{AGPC4} into the gauge constraint \eqref{reggauge} (but now leaving $\alpha_{2 n + 1}$ as undetermined coefficients), we obtain the following $2$ relations
\begin{align}
    &i b_{1} + \alpha_{1}(b_{1}^2 + b_{2}^2 + \mu^2) + \alpha_{3}b_{3}b_{2} = 0\label{eq-1} \\
    &\alpha_{2 L - 1}b_{2 L}b_{2 L + 1} + \alpha_{2 L + 1}(b_{2 L + 1}^2 + b_{2 L + 2}^2 + \mu^2) = 0\label{eq0}
\end{align}
And the following set of $M-1$ relations
\begin{align}
    \alpha_{2 k - 1}&b_{2 k}b_{2 k + 1} + \alpha_{2 k + 1}(b^2_{2 k + 2} + b_{2 k + 1}^2 + \mu^2)\notag \\ + &\alpha_{2 k + 3}b_{2 k + 3}b_{2 k + 2} = 0  \;\;\forall k \in \{1, M - 1\} \label{eq1}
\end{align}
This set of equations can be written in the matrix form as follows
\begin{widetext}
\begin{gather}
 \begin{bmatrix} 
 b_{1}^2 + b_{2}^2 + \mu^2 & b_2 b_3 & 0 & 0 &  \cdots & 0 \\
 b_2 b_3 & b_{3}^2 + b_{4}^2 + \mu^2 & b_4 b_5 & 0 &  \cdots & 0 \\
 0 & b_4 b_5 & b_{5}^2 + b_{6}^2 + \mu^2 & b_6 b_7 & \cdots & 0 \\
 \vdots & \vdots & \vdots & \vdots & \ddots & \vdots\\
 0 & 0 & 0 & \cdots & b_{2 L}b_{2 L + 1} & b_{2 L + 1}^2 + b_{2 L + 2}^2 + \mu^2
 \end{bmatrix}
  \begin{bmatrix}
   \alpha_{1} \\
   \alpha_{3} \\
   \alpha_{5} \\
   \vdots \\
   \alpha_{2 L + 1}
   \end{bmatrix} 
   =
   \begin{bmatrix}
   -i b_1 \\
   0 \\
   0 \\
   \vdots \\
   0
   \end{bmatrix}\label{mateqn}
\end{gather}
\end{widetext}
Once the Lanczos coefficients $b_n$ have been evaluated numerically by implementing the Lanczos algorithm, the coefficients of $\mathcal{O}_{2 n + 1}$ can be determined by solving the matrix equation given above.

\section{Analytical Examples}
In this section, we evaluate some explicit expressions for the AGP using the expression \eqref{AGPC2}. The full steps of the evaluation are covered in Appendix \ref{appB}. In the models, we require analytic expressions of the autocorrelation function. So we are naturally restricted to small integrable systems. In Appendix\,\ref{appD}, we derive an approximate analytical expression for the AGP norm of the Ising chain at criticality.
\subsection{A $2$-level system}
We consider the following model \cite{PhysRevLett.109.115703} (demonstrating the Landau-Zener transition \cite{zener1932non}), which is also the simplest model supporting the Kibble-Zurek mechanism \cite{kibble1976topology, kibble1980some, zurek1996cosmological}. This model is described by the simple Hamiltonian
\begin{align}
	H(\lambda) = \lambda \sigma^{z} + \Delta \sigma^{x}\,.\label{2levsys}
\end{align}
The autocorrelation function for this system can be easily evaluated, say for the operator $\sigma^{z}$
\begin{align}
	\mathcal{C}(t) &= \frac{\tr{e^{i H t}\sigma^{z}e^{-i H t}\sigma^{z}}}{\tr{\mathbf{1}_{2 \times 2}}}\notag \\ &= \frac{\left(\lambda ^2+\Delta ^2 \cos \left(2 t \sqrt{\Delta ^2+\lambda ^2}\right)\right)}{(\Delta ^2+\lambda ^2)}\,.
\end{align}
Inserting this expression in \eqref{AGPC2}, we obtain the following expression
\begin{align}
	\vert \vert A_\lambda \vert \vert^{2} = \frac{4 \Delta ^2}{\left(4 \left(\Delta ^2+\lambda ^2\right)+\mu ^2\right)^2} \xrightarrow[\mu \rightarrow 0]{} \frac{\Delta ^2}{4(\Delta ^2+\lambda ^2)^{2}}\label{LZn}
\end{align}
For this system, we can take the limit $\mu \rightarrow 0$ without any issues because it does not exhibit any degeneracies. We compare this result with the AGP operator derived in \cite{PhysRevLett.109.115703}. 
Using the Krylov approach, one can use \eqref{AGPC1} to derive the AGP operator. This calculation is described in detail in Appendix \ref{appB2l}. The operator thus obtained is written below.
\begin{align} 
	A_\lambda = -\frac{1}{2}\frac{\Delta}{\Delta^{2} + \lambda^{2}}\sigma^{y}
\end{align}
The norm calculated from this AGP operator, using \eqref{agporg}, can be shown to be
\begin{align}
	\vert \vert A_\lambda \vert \vert^{2} &= \frac{1}{\mathcal{D}}\sum_{m \neq n}\vert\bra{m}A_{\lambda}\ket{n}\vert^{2} \notag \\ &=  \frac{\Delta^{2}}{4(\Delta^{2} + \lambda^{2})^{2}}
\end{align}
which is the result obtained in \eqref{LZn}, happily.

\subsection{A 2-qubit system}
We consider the following simple $2$-qubit system \cite{petiziol} given by the Hamiltonian
\begin{align}
	H(\lambda) = - \left(\sigma^{1}_{x}\sigma^{2}_{x} + \sigma^{1}_{z}\sigma^{2}_{z} \right) - \varepsilon (1 - \lambda)\left( \sigma^{1}_{z} + \sigma^{2}_{z} \right)\,.\label{2q}
\end{align}
Evaluating the autocorrelation function $\mathcal{C}(t)$ (starting with $\mathcal{O}_0 = \frac{1}{\sqrt{2}}(\sigma^{1}_{z} + \sigma^{2}_{z})$) is straightforward and gives the following result
\begin{align}
	\mathcal{C}(t) &= \frac{\tr{e^{i H t}\mathcal{O}_0 e^{-i H t}\mathcal{O}_0}}{\tr{\mathbf{I}_{2 \times 2}}}\notag \\ &= \frac{4 \varepsilon^2 ( 1 - \lambda)^2 + \cos(2 t \sqrt{1 + 4 \varepsilon^2(1-\lambda)^2})}{1 + 4 \varepsilon^2 (1 -\lambda)^2}\,. \label{auc2q}
\end{align}
From this expression, we evaluate the AGP norm using \eqref{AGPC2} 
\begin{align}
	\vert \vert A_\lambda \vert \vert^2 =  \frac{8}{\left(16 \varepsilon^2 (1-\lambda)^2+4\right)^2}\,.\label{2qn}
\end{align}
Note that this Hamiltonian \eqref{2q} is non-degenerate for real $\varepsilon, \lambda$. Therefore, it is possible to take the limit $\mu \rightarrow 0$ in the AGP norm and operator without causing a divergence. 

We employ the Lanczos algorithm in order to derive the form of the AGP operator, using \eqref{AGPC1}. The result is (derivation in Appendix \ref{appB2q})
\begin{align}
	A_\lambda = -i\frac{\sqrt{2}}{4(1 + 4 \varepsilon^2 (1-\lambda)^2}\left(\sigma^{1}_{x}\sigma^{2}_{y} + \sigma^{1}_{y}\sigma^{2}_{x} \right)\,.
\end{align}
This is not the result discussed in \cite{petiziol}. This is because in \cite{petiziol}, the AGP operator is derived with respect to the unnormalized $\partial_{\lambda}H$. The normalization factor $\sqrt{2}\varepsilon$ can be reinserted into the AGP operator to obtain the result in \cite{petiziol}. This implies multiplying the AGP norm \eqref{2qn} by $2\varepsilon^2$.
\subsection{A 4-body system}
\begin{figure}[htbp]
	\centering
	\includegraphics[width=0.5\textwidth]{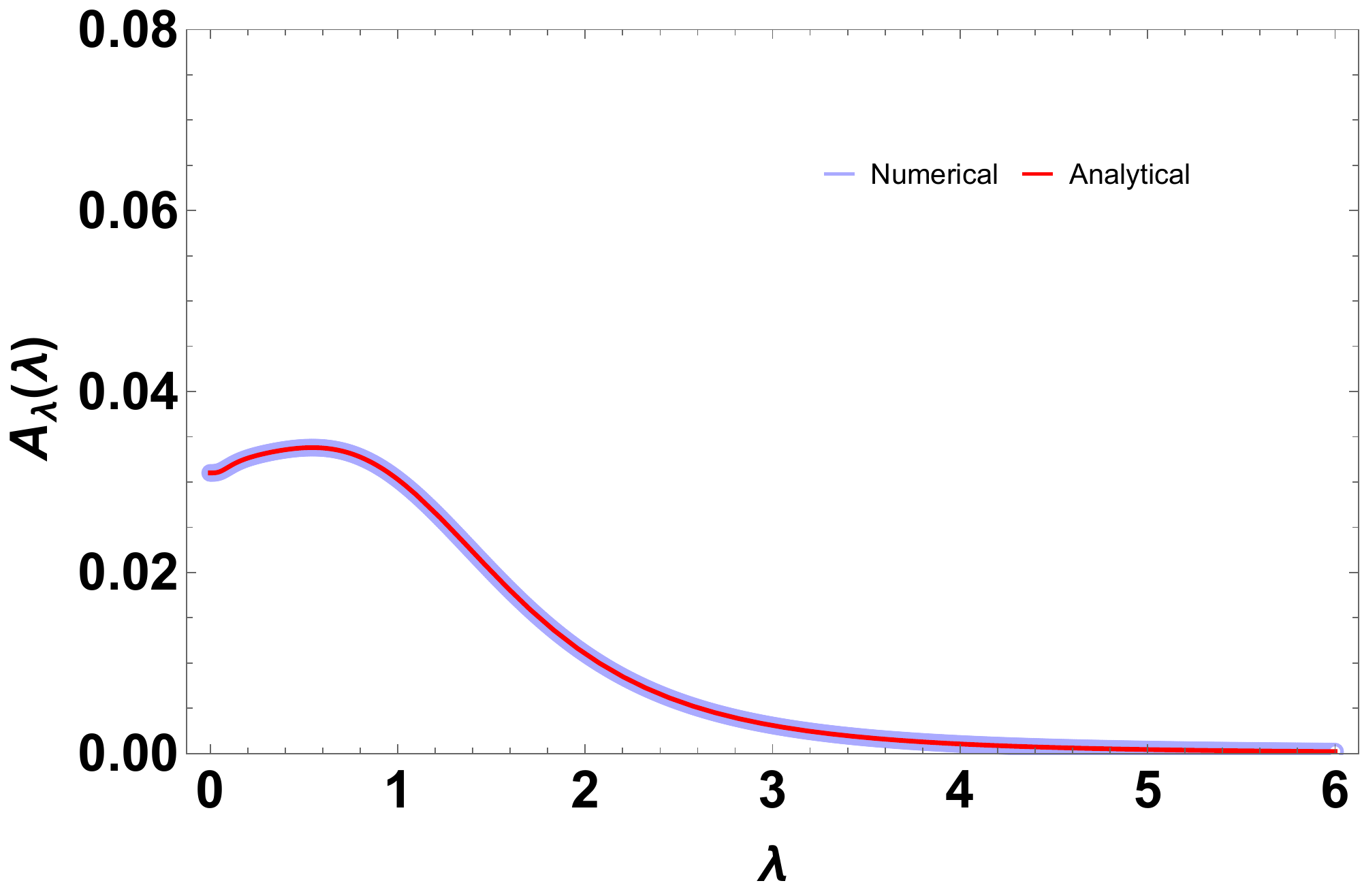}
	\caption{The AGP norm computed numerically for the Hamiltonian \eqref{4body} and compared with the norm of the analytical expression \eqref{AGP4b}. It is clear that the expression exactly captures the AGP, as per expectations. } \label{fig:highmuhnn}
\end{figure}

In this section, we consider a simple $4$-body system described by the Hamiltonian
\begin{align}
	H = \sum_{i = 0}^{3}\sigma^{x}_{i}\sigma^{x}_{i + 1} + \lambda(\sigma^z_{1} + \sigma^{z}_2)\,.\label{4body}
\end{align}
The Krylov basis operators and the Lanczos coefficients are given in Appendix \ref{appB4}. The initial operator is chosen as $\partial_\lambda H = \sigma^{z}_1 + \sigma^{z}_2$. The regulator is taken to be $\mu = L 2^{-L}$, with the system size $L = 4$ in this case. We evaluate the AGP operator 
\begin{align}
	A_\lambda &= c_{1} (\sigma^{z}_1 + \sigma^{z}_2) + c_{2}\left( \sigma^{x}_0 \sigma^{y}_1 + \sigma^{y}_2 \sigma^{x}_3\right) \notag \\ &+ c_{3}\left(\sigma^{x}_1 \sigma^{y}_2 + \sigma^{y}_1 \sigma^{x}_2\right) + c_{4}\left(\sigma^{x}_0 \sigma^{z}_1 \sigma^{y}_2 + \sigma^{y}_1 \sigma^{z}_2 \sigma^{x}_3 \right)\label{AGP4b}
\end{align}
where
\begin{widetext}
	\begin{align}
		c_{1} &= -\frac{i \left(\frac{3 \sqrt{2} \lambda  \left(9 \lambda ^6-184 \lambda ^4+512 \lambda ^2+1024\right) \left(\lambda ^2 \left(9 \mu ^2+272\right)+36 \lambda ^4+32 \mu ^2\right)}{\left(5 \lambda ^2+4\right) \left(9 \lambda ^2+32\right) \left| 16-9 \lambda ^2\right| }+2 \left(8 \left(\lambda ^2+2\right) \mu ^2+\frac{16 \lambda ^2 \left(9 \lambda ^4+50 \lambda ^2+256\right)}{9 \lambda ^2+32}+\mu ^4\right)\right)}{8 \left(3 \lambda ^2+4\right) \mu ^4+16 \left(3 \lambda ^2+4\right)^2 \mu ^2+64 \lambda ^2 \left(4 \lambda ^4+13 \lambda ^2+32\right)+\mu ^6} \\
		c_{2} &= -\frac{12 \sqrt{2} \lambda ^2 \left(\lambda ^2 \left(5 \mu ^2+32\right)+2 \lambda ^4+4 \left(\mu ^2+16\right)\right)}{\left(5 \lambda ^2+4\right) \left(8 \left(3 \lambda ^2+4\right) \mu ^4+16 \left(3 \lambda ^2+4\right)^2 \mu ^2+64 \lambda ^2 \left(4 \lambda ^4+13 \lambda ^2+32\right)+\mu ^6\right)} \\
		c_{3} &= \frac{8 \sqrt{2} \lambda ^2 \left(27 \lambda ^4+16 \lambda ^2-64\right)}{\left(5 \lambda ^2+4\right) \left(8 \left(3 \lambda ^2+4\right) \mu ^4+16 \left(3 \lambda ^2+4\right)^2 \mu ^2+64 \lambda ^2 \left(4 \lambda ^4+13 \lambda ^2+32\right)+\mu ^6\right)} \\
		c_{4} &= -\frac{16 \sqrt{2} \lambda  \left(7 \lambda ^2+\mu ^2\right)}{8 \left(3 \lambda ^2+4\right) \mu ^4+16 \left(3 \lambda ^2+4\right)^2 \mu ^2+64 \lambda ^2 \left(4 \lambda ^4+13 \lambda ^2+32\right)+\mu ^6}
	\end{align}
\end{widetext}
The AGP norm evaluated from the expression \eqref{AGP4b} is plotted in Fig.\,\ref{fig:highmuhnn} and compared with the numerically evaluated AGP norm directly from the Hamiltonian \eqref{4body}.

\section{Approximating the AGP}
We have described a systematic procedure to evaluate the AGP operator. However, evaluating the full Krylov basis and all the Lanczos coefficients analytically is a painful process. Even numerically, evaluating the full AGP can prove to be non-trivial via the Lanczos approach, especially if there exists an exponentially large Krylov basis implying an exponentially large number of equations \eqref{eq1} that need to be solved to evaluate the AGP norm. Therefore, it is not desirable to solve the full set of equations in \eqref{eq1} in order to determine the AGP.

In view of the difficulty in evaluating the full Krylov basis, we consider approximating the AGP by truncating the series at \eqref{AGPC1} at some $N < M$. We terminate the series at $n = N < M$ and proceed to solve \eqref{eq1} with a reduced set of equations. Consequently, the dimensions of the matrix \eqref{mateqn} will be lesser and therefore easier to solve the matrix equation for the reduced vector $\{\alpha_{1}, \alpha_2, \alpha_3,\dots,\alpha_{2 N + 1}\}$. The expression for the AGP operator is then given as
\begin{align}
	A_\lambda = \sum_{n = 0}^{N < M}\alpha_{2 n + 1}\mathcal{O}_{2 n + 1}\label{AGPCtrunc}
\end{align}
It is easier to evaluate the initial few Krylov vectors as compared to the ones at higher index values analytically. An explicit form of the AGP operator can be approximated\footnote{As we shall see, it is not a very good approximation} by considering a small value of $N$. The expression obtained for the AGP norm $\vert \vert A_\lambda \vert \vert^2 = - \sum_{n = 0}^{N < M}\alpha_{2 n + 1}^2$ using the truncated series can be compared with the exact numerical result to determine the number of terms that need to be included to get a reasonably approximate AGP operator. 

\subsection{The Variational Approach}

In \cite{selspol}, a variational approach for evaluating the AGP operator was proposed. We now discuss the effect of truncation with respect to this approach. A similar discussion, with close results, can be found in \cite{hatomura2021controlling}. For the regulated AGP, the variational approach involves minimizing the action, defined as 
\begin{align}
    S = \tr{G_\lambda^2} = \tr{(i\partial_\lambda H + [H,A_\lambda])^2 + \mu^2 A_\lambda^2}
\end{align}
with respect to an appropriately chosen set of free parameters in the AGP operator $A_\lambda$. The approximate AGP operator is found by solving for the following Euler-Lagrange equation
\begin{align}
    \frac{\delta S(A_\lambda)}{\delta A_\lambda} = 0
\end{align}
We observe that choosing the $A_\lambda$ to be given by the expression \eqref{AGPCtrunc} gives us the following expression for the action
\begin{align}
    S &= (i + \alpha_1 b_1)^2 + \sum_{k=1}^{N}(\alpha_{2 k - 1}b_{2 k} + \alpha_{2 k + 1}b_{2 k + 1})^2 \notag \\ &+ \alpha_{2 N + 1}^2 b_{2 N + 2}^2 + \mu^2\sum_{k = 0}^{N}\alpha_{2 k + 1}
\end{align}
It is straightforward to see that extremizing the action $S$ with respect to the free parameters $\alpha_{2 k + 1}$ gives back the equations \eqref{eq-1}-\eqref{eq1}, with $N$ instead of $L$. This is expected since any additional Krylov basis operators (added to $A_\lambda$) are orthonormal to the terms already included. Therefore, there is no ``weight adjustment'' effect (or a kind of ``back-reaction'') on the coefficients equations determining the coefficients $\alpha_{2 k + 1}$. It is important to note that adding more terms \textit{does affect} the explicit expressions of the $\alpha_{2 k + 1}$ (in terms of the Lanczos coefficients) already determined as they are solutions of a coupled set of linear equations.

\section{Numerical Results}

We study this truncation process in three types of systems: \textit{integrable}, \textit{weakly chaotic}\footnote{There are various definitions of \textit{Weak Chaos}. We choose such systems which are (classically) fully integrable except for a finite number of unstable saddle points in their phase space.}, and \textit{strongly chaotic}\footnote{We choose a system that is chaotic with respect to level spacing statistics.}. We find that due to the successively (nearly exponentially) increasing Krylov space dimensions for these systems, the truncation method shows decreasing levels of success in capturing the basic features of the AGP.

\subsection{Integrable Free System}
\begin{figure}[htbp]
	\subfigure[]{\includegraphics[height=5.7cm,width=1\linewidth]{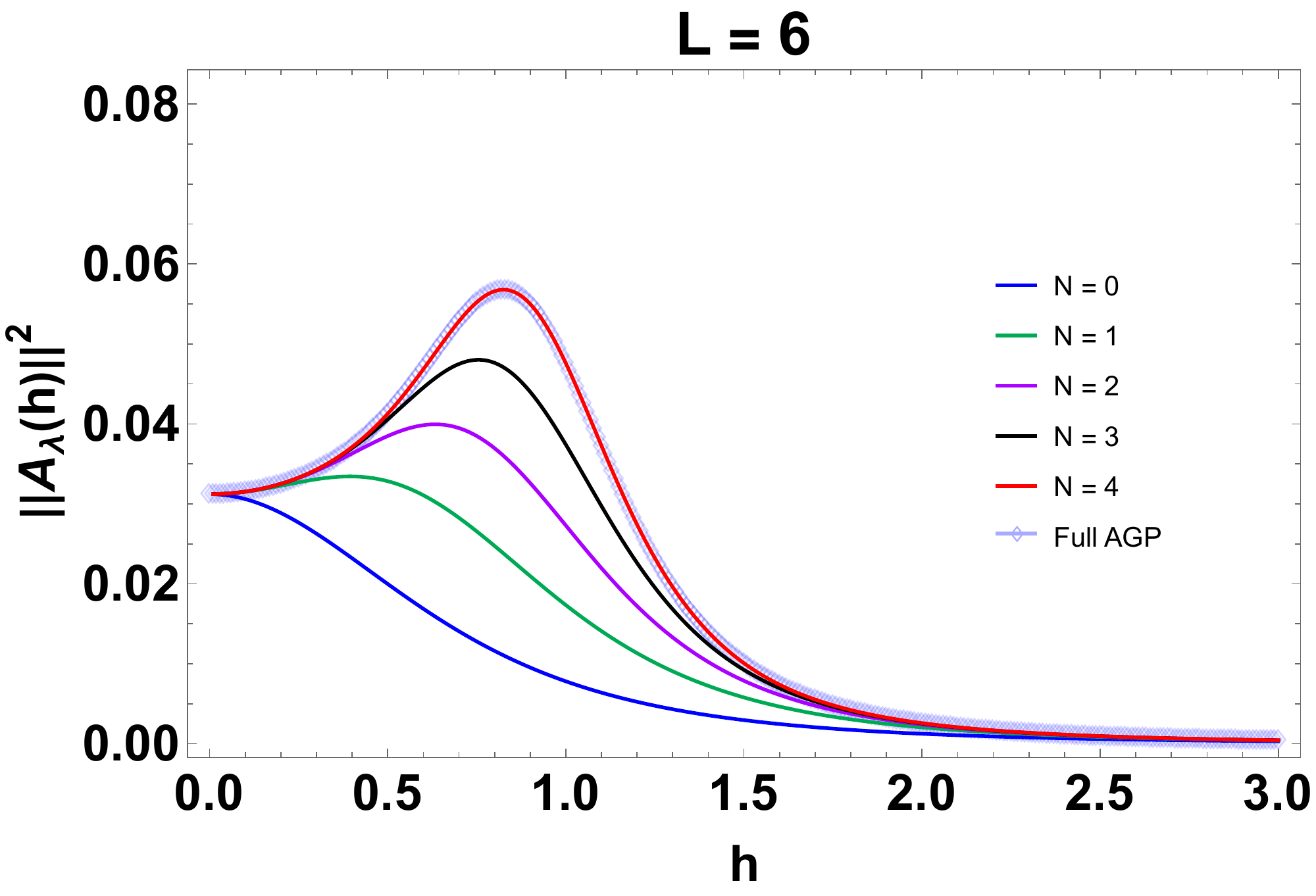}\label{fig:AGPIntIsing06}}
	\hfill
	\subfigure[]{\includegraphics[height=5.7cm,width=1\linewidth]{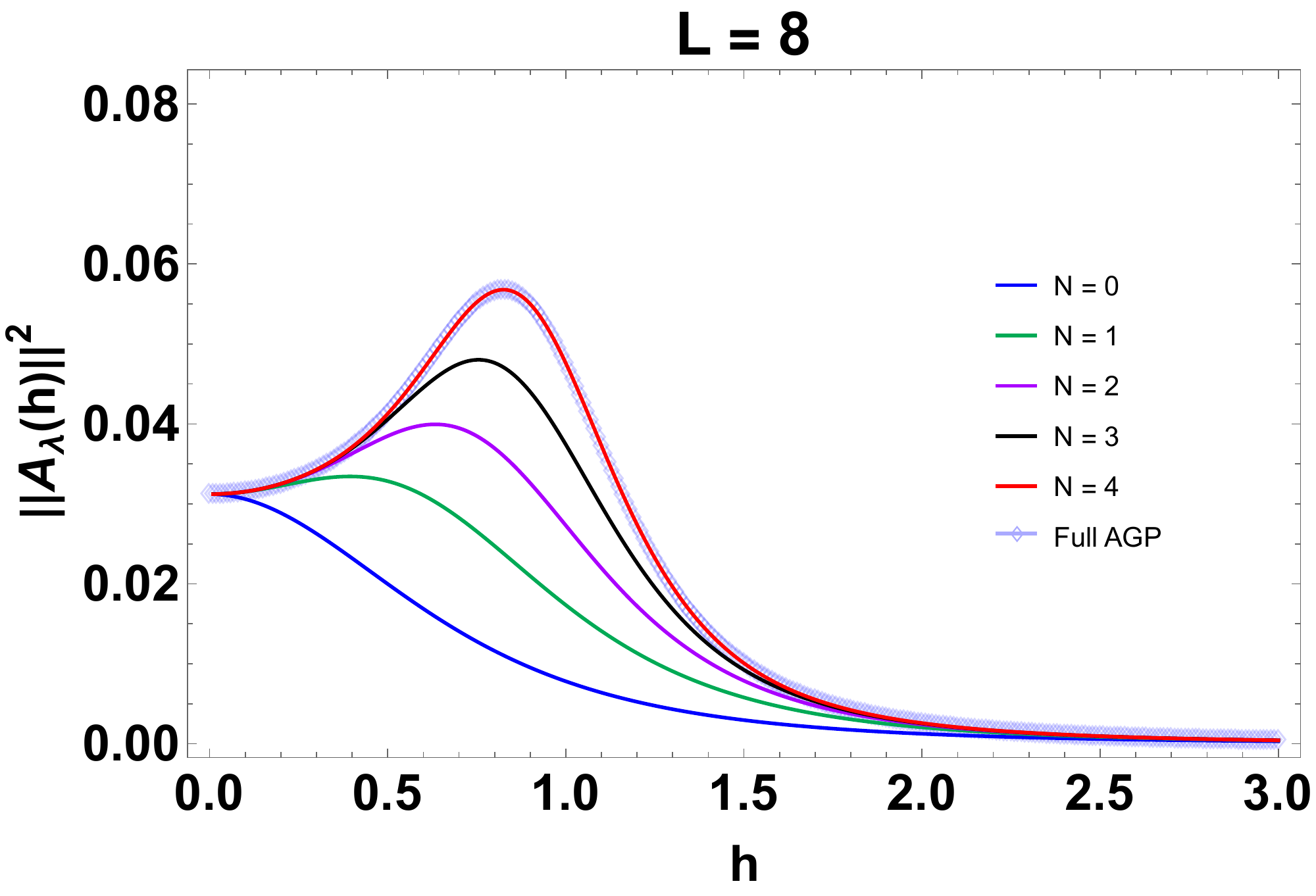}\label{fig:AGPIntIsing08}}
	\hfill
	\subfigure[]{\includegraphics[height=5.7cm,width=1\linewidth]{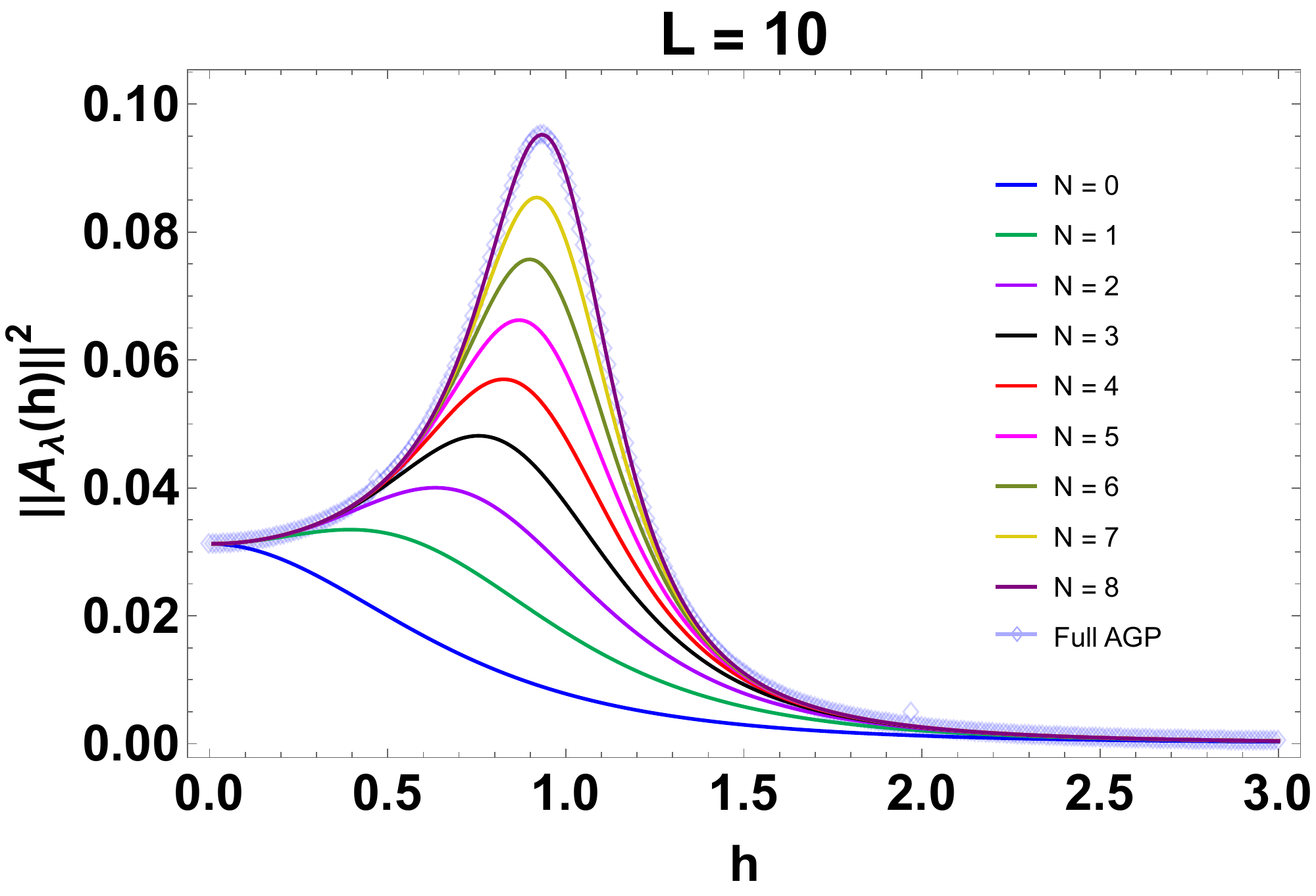}\label{fig:AGPIntIsing10}}
	\caption{The AGP norm is compared by direct evaluation \eqref{AGPnorm5} and by solving \eqref{mateqn} for different cutoffs. It can be seen that length of the vector $\vec{\alpha}$ in \eqref{mateqn} is (a) $M = 4$ for system size $6$,(b) $M = 6$ for system size $8$ and (c) $M = 8$ for system size $10$. The phase transition in the model at $h = 1$ is also picked by the AGP, allowing for finite-size effects. The regulator is chosen to be $\mu = L  2^{-L}$.}\label{IntIsingAGP}
\end{figure}

We consider the integrable (non-interacting) Ising model with periodic boundary conditions, given by the Hamiltonian below
\begin{align}
	H = \sum_{i = 1}^{L}\sigma^{z}_{i}\sigma^{z}_{i+1} + h\sigma^{x}_{i}\label{IntegrableIsingModel}
\end{align}
with the identification $\sigma^{z}_{N + 1} = \sigma^{z}_1$. The analytical expression for the integrable Ising model AGP was derived in \cite{PhysRevLett.109.115703}. Due to the integrable (non-interacting) nature of the model, the Krylov space is highly restricted. This is reflected by computing the AGP norm exactly by solving the matrix equation \eqref{mateqn} for small $M$. The results are given in Fig.\,\ref{IntIsingAGP} for system size $L = 6, 8$ and $10$. 

\begin{figure}[htbp]
	\includegraphics[height=5.7cm,width=\linewidth]{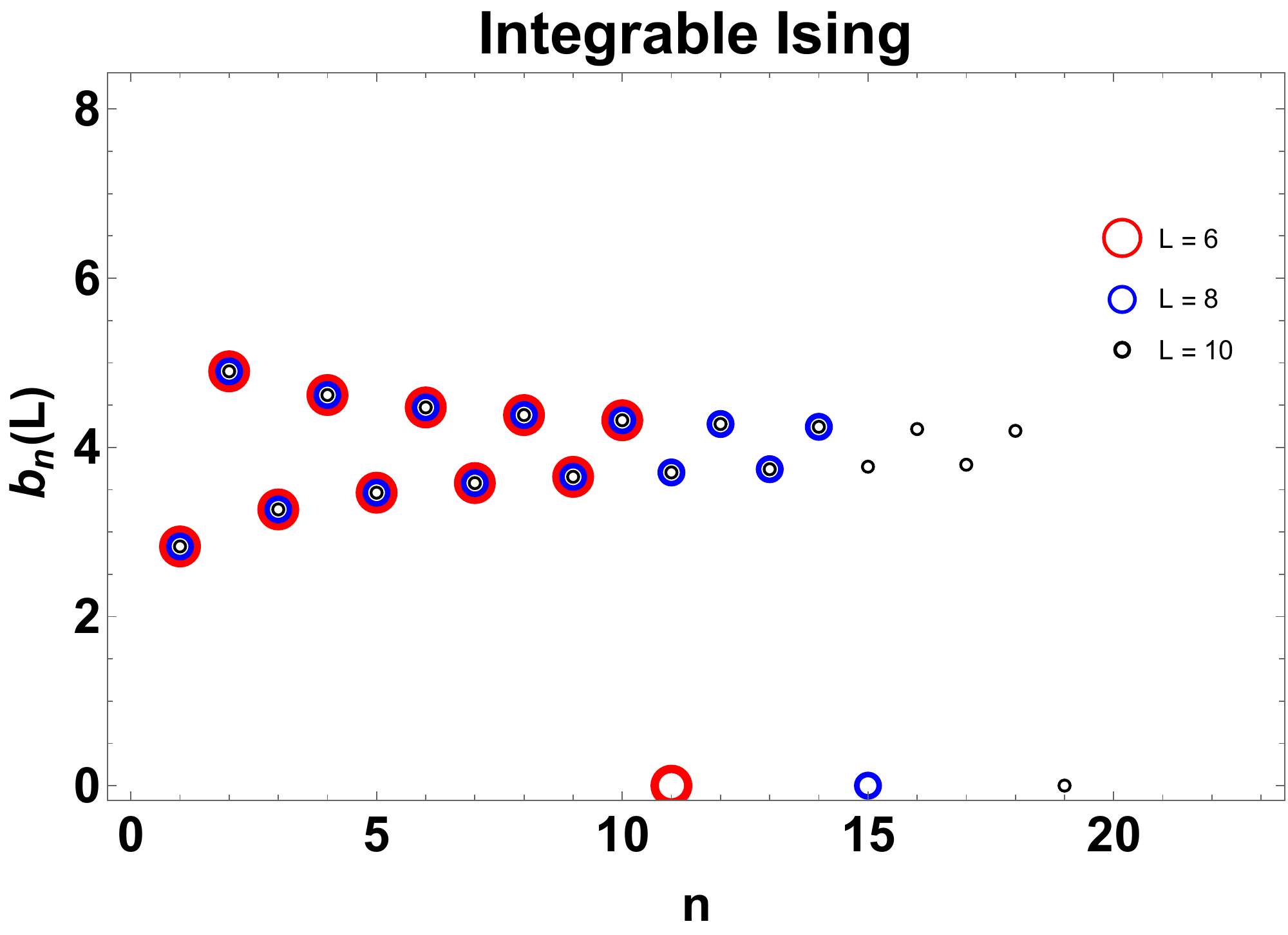}
	\caption{The Lanczos coefficients for system size $6$, $8$ and $10$ and $h = 1$. The Krylov space is highly restricted, as only the first few Lanczos coefficients are non-zero. A nice feature of the model is that the Lanczos coefficients (and consequently the \textit{coefficients} of the terms in the AGP operator) are independent of the system. The system size contributes by affecting the total number of terms in the expression of the AGP operator. }\label{fig:AGPIntIsingLanczos}
\end{figure}

The finiteness of the Krylov space can be measured by evaluating the Lanczos coefficients numerically. These are shown in Fig.\,\ref{fig:AGPIntIsingLanczos}. It is observed that the Krylov space is much smaller than $\mathcal{D}^2 - \mathcal{D} + 1$, where $\mathcal{D} = 2^{L}$ for system size $L$. This is a sign of the integrable nature of the system. It is interesting to note that the Lanczos coefficients do not depend on system size. The only contribution of the system size to the AGP is in terms of the number of Krylov basis operators required to describe the AGP. The coefficients of these operators (in the expression for AGP operator) are independent of the system size itself.

\subsection{Weakly chaotic system} 
To study the truncation method in a weakly chaotic system, we turn our attention to the Lipkin-Meshkov-Glick (LMG) model \cite{LIPKIN1965188}. This model is known to be weakly chaotic in the sense that it displays saddle-dominated scrambling. The classical limit of this model is integrable. However, it possesses one unstable saddle point, in whose vicinity orbits are chaotic. The effect of the saddle point has been studied from the perspective of Out-of-Time-Ordered-Correlator (OTOC)\cite{Xu:2019lhc} and Krylov complexity \cite{Bhattacharjee:2022vlt}. It was found that K-complexity is hypersensitive to the presence of unstable saddle points. The LMG model Hamiltonian is given below
\begin{align}
	H = \hat{X} + 2 J \hat{Z}^2\label{LMGHam}
\end{align}
where $\{\hat{X},\hat{Y},\hat{Z}\}$ follow the $SU(2)$ algebra. We study this model for spin realizations $S = 10$ and $30$ and compare the numerically evaluated AGP norm \eqref{agporg2} with the one calculated from the Lanczos coefficients by taking a finite number of Krylov vectors to describe the AGP operator. The adiabatic deformation parameter $\lambda$ is taken to be $2 J$ in this case. The intial operator is $\partial_\lambda H = \hat{Z}^2$. We begin by normalizing this operator and performing the Lanczos algorithm. 

The first few operators in the Krylov basis are given below
\begin{align}
	\mathcal{O}_{0} &= \frac{1}{z}\hat{Z}^2 \\
	\mathcal{O}_{1} &= -\frac{1}{z b_{1}}\left(\hat{Z}\hat{Y} + \hat{Y}\hat{Z}\right)\\
	\mathcal{O}_{2} &= \frac{2}{z b_{1} b_{2}}\left(\hat{Y}^2 - \hat{Z}^2 + 2 J (\hat{X}\hat{Z}^2 + \hat{Z}^2 \hat{X}^2 + 2 \hat{Z} \hat{X} \hat{Z} \right)\notag \\ &- \frac{b_{1}}{b_{2} z}\hat{Z}^2 \\
	\mathcal{O}_{3} &= \frac{2}{z b_{1} b_{2} b_{3}}\Big(-2 z b_{1}\mathcal{O}_{1} + 2 J z b_{1} \{\hat{X},\mathcal{O}_{1} \}\notag\\ &- 4 J^2 z b_{1}\{\hat{Z},\mathcal{O}_{1}\} - 8 J^2 z b_{1}\hat{Z}\mathcal{O}_{1}\hat{Z} \notag \\ &-2J\left(\hat{Y}\hat{Z}\hat{X} + \hat{Z}\hat{X}\hat{Y} + \hat{Y}\hat{X}\hat{Z} + \hat{X}\hat{Z}\hat{Y} \right)\notag\\ &- 4 J (\hat{Z}\hat{X}\hat{Y} + \hat{Y}\hat{X}\hat{Z}) \Big) + \frac{1}{z b_3}\left(\frac{b_2}{b_1} - \frac{b_1}{b_2} \right)\mathcal{O}_{1}
\end{align}
Here the norm of $\mathcal{O}_{0}$ is given by $z = \sqrt{|\hat{Z}^2|} = \frac{\sqrt{S (S+1) \left(3 S^2+3 S-1\right)}}{\sqrt{15} S^2}$ for arbitary spin $S$. To evaluate the regulated AGP norm, we pick the regulator $\mu = (2 S + 1)2^{-(2 S + 1)}$.

The operators become very complicated very quickly. The Lanczos coefficients are hard to evaluate for general spin $S$. The coefficients are dependent on both $S$ and $J$ (i.e. of $O(J)$). For the cases we study numerically ($S = 10$ and $S = 30$), the first few Lanczos coefficients are listed in Table.\,\ref{table}.

\begin{table}[h]
	\begin{tabular}{ |c|c|c| } 
		\hline
		$b_{n}$ & $S = 10$ & $S = 30$ \\ 
		\hline 
		\hline
		$n = 1$ & $0.12$ & $0.038$ \\
		\hline
		$n = 2$ & $\sqrt{0.074 J^2+0.027}$ & $\sqrt{0.0079 J^2+0.0030}$ \\
		\hline
		$n = 3$ & $J\sqrt{\frac{0.021 J^2+0.053}{J^2+0.36}}$ & $J\sqrt{\frac{0.0023 J^2+0.0059}{J^2+0.38}}$ \\
		\hline
	\end{tabular}
	\caption{Listing the first three Lanczos coefficients for $S = 10$ and $S = 30$ realizations of the LMG model}\label{table}
\end{table}

\begin{figure}[htbp]
	\subfigure[]{\includegraphics[height=5.7cm,width=1\linewidth]{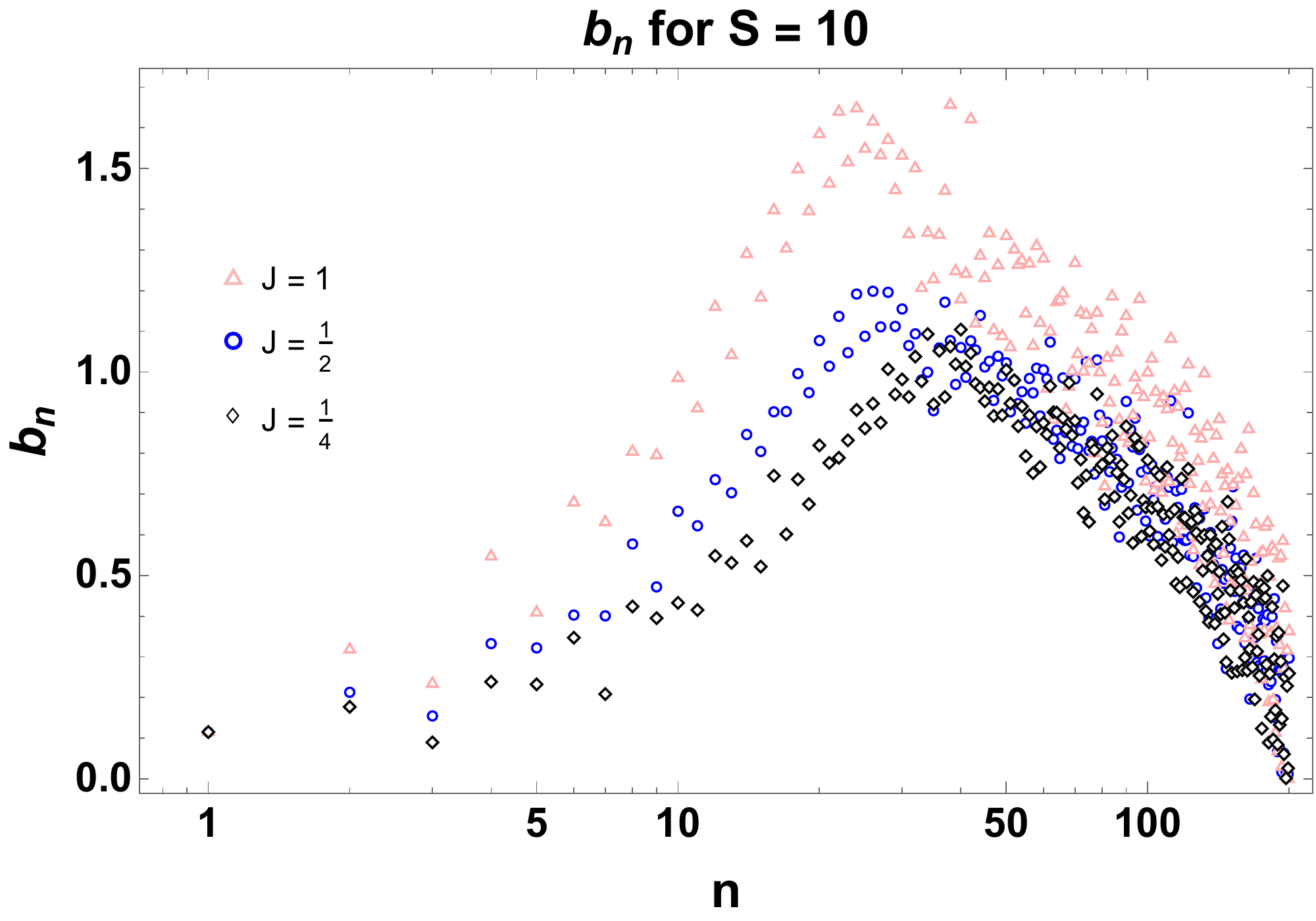}\label{fig:LMGS10Lanczos}}
	\hfill
	\subfigure[]{\includegraphics[height=5.7cm,width=1\linewidth]{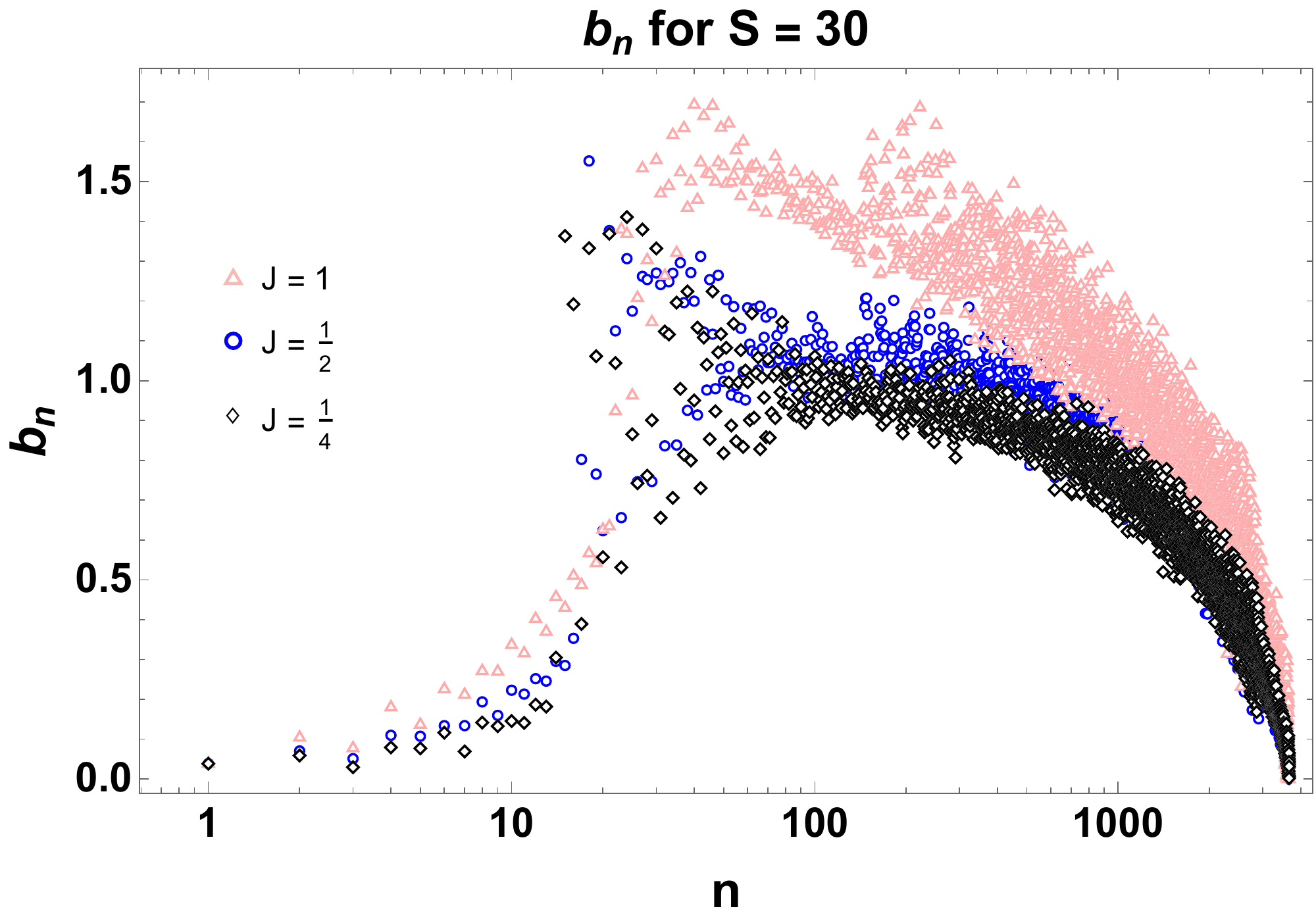}\label{fig:LMGS30Lanczos}}
	\caption{The full set of Lanczos coefficients are computed for $J = \{\frac{1}{4}, \frac{1}{2}, 1\}$ for spin realisations (a) $S = 10$ and (b) $S = 30$. It can be observed that the number of Lanczos coefficients for (a) $S = 10$ is $2 M + 1 = 201$, which gives us $M = 100$. Similarly for (b) $S = 30$, we have $2 M + 1 = 3659$, which gives us $M = 1829$. Note that both satisfy the relation $K \leq \mathcal{D}^2 - \mathcal{D} + 1$, where $\mathcal{D} = 21$ for $S = 10$ and $\mathcal{D} = 61$ for $S = 30$. }\label{LMGLanczos}
\end{figure}

\begin{figure}[htbp]
	\subfigure[]{\includegraphics[height=5.7cm,width=1\linewidth]{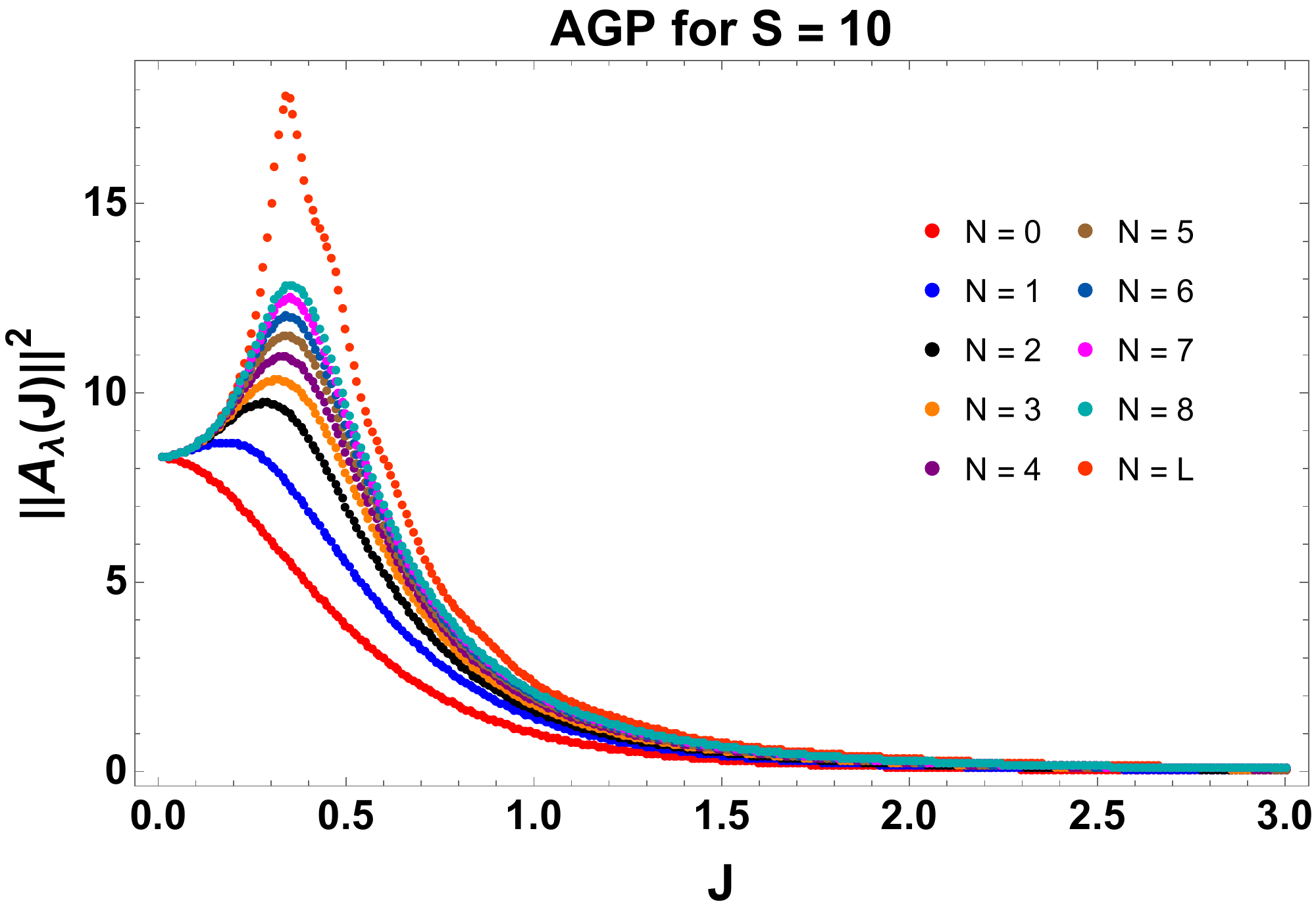}\label{fig:LMGS10}}
	\hfill
	\subfigure[]{\includegraphics[height=5.7cm,width=1\linewidth]{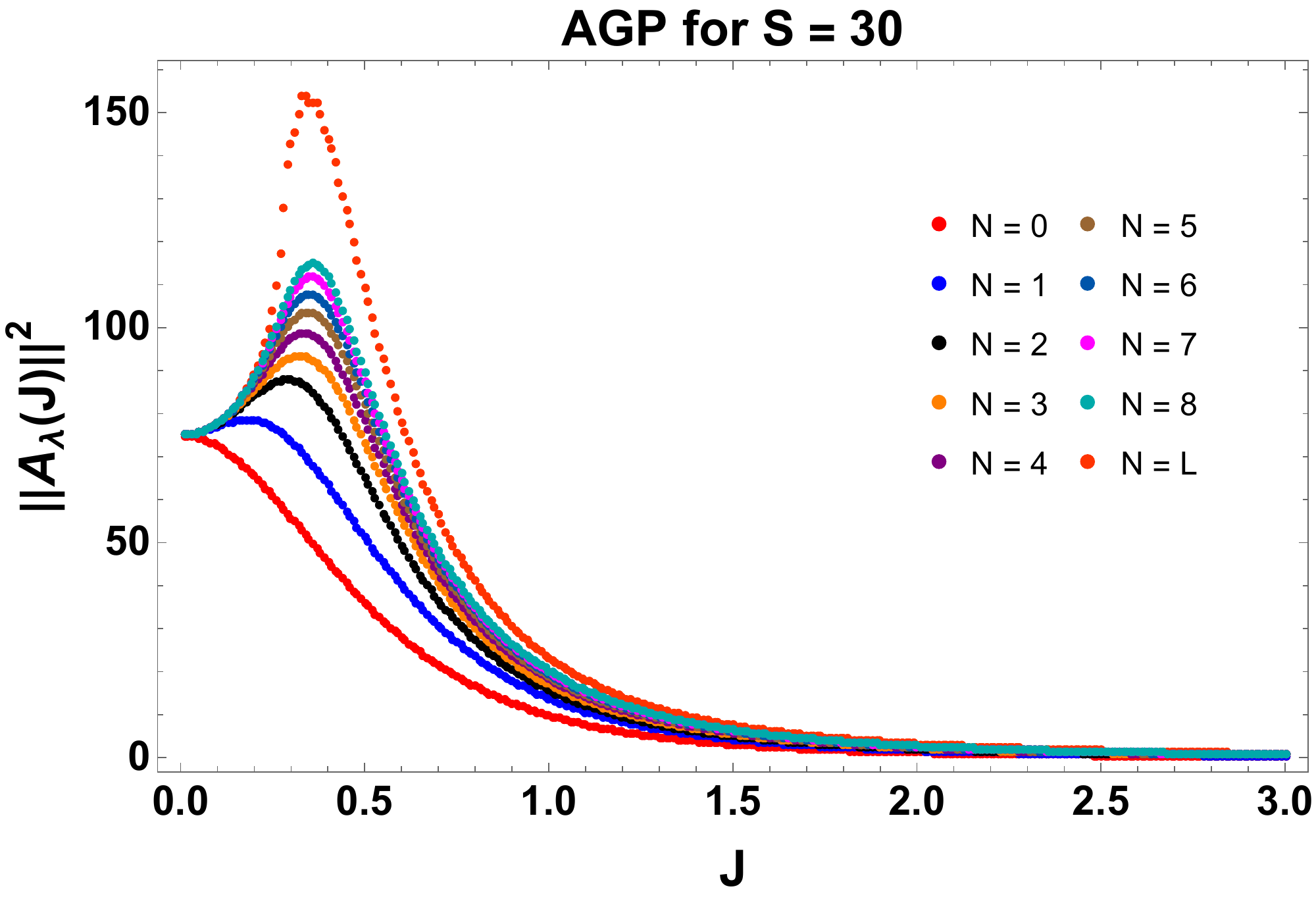}\label{fig:LMGS30}}
	\caption{(a) The full AGP norm ($N = L$) and the truncated AGP norm evaluated at orders $N = 0,\dots,8$ for the LMG model at a spin realisation $S = 10$. Here the adiabatic deformation parameter is $\lambda = 2 J$. (b) The full AGP ($N = L$) norm and the truncated AGP norm ($N = 0,\dots,8$) for $S = 30$ realization of the LMG model. }\label{fig:LMGAGP}
\end{figure}

We demonstrate numerically that the qualitative behavior of the AGP is captured by using \eqref{mateqn} with the first few Lanczos coefficients. The matrix in \eqref{mateqn} is constructed by cutting off at successive lengths $N$, and the AGP norm arising out of that\footnote{We shall refer to this as \textit{truncated AGP}} is evaluated. This is done by considering \eqref{mateqn} by terminating the matrix at $2 N + 1 < 2 M + 1$ and solving the matrix equation to evaluate $\alpha_{2 k + 1}$ where $0 \leq k \leq N < M$. We perform this calculation for the two spin realizations of the LMG model ($S = 10$ and $S = 30$). The numerical results for the truncated AGP and the full AGP are given in Fig.\,\ref{fig:LMGS10} for $S = 10$ and Fig.\,\ref{fig:LMGS30} for $S = 30$.

One can find $M$ by evaluating the full set of Lanczos coefficients. The total number of Lanczos coefficients is the extent of the Krylov space. The full Krylov space is required to describe the AGP exactly. In Figs.\,\ref{fig:LMGS10Lanczos} and \ref{fig:LMGS30Lanczos}, the full set of Lanczos coefficients are computed for the $S = 10$ and $S = 30$. The Krylov space ends once the Lanczos coefficients become $0$.

Note that there is a phase transition in the classical LMG model. The system transitions from an integrable to a chaotic phase at $J = \frac{1}{2}$. This can be observed via a classical calculation \cite{Bhattacharjee:2022vlt}, which corresponds to taking $S \rightarrow \infty$. The AGP (evaluated for finite $S$) reflects this phase transition since the AGP peaks close to $J = \frac{1}{2}$, as seen in Fig.\,\ref{fig:LMGAGP}.

\subsection{Integrable Interacting model}
We consider the $XXZ$ model with open boundary conditions \cite{PhysRevX.10.041017}. The Hamiltonian is the following
\begin{align}
	H = \sum_{i = 1}^{L-1}\left(\sigma^{x}_{i+1}\sigma^{x}_{i} + \sigma^{y}_{i+1}\sigma^{y}_{i}\right) + \Delta \sum_{i = 1}^{L-1}\sigma^{z}_{i+1}\sigma^{z}_{i}\label{HamXXZ}
\end{align}
This system is Bethe ansatz solvable \cite{Franchini:2016cxs}. However, the spectrum is not known, and thus the exact AGP is not available. Therefore, despite being integrable, it is different from the integrable non-interacting Ising model studied in a previous section.

\begin{figure}[htbp]
	\subfigure[]{\includegraphics[height=5.7cm,width=1\linewidth]{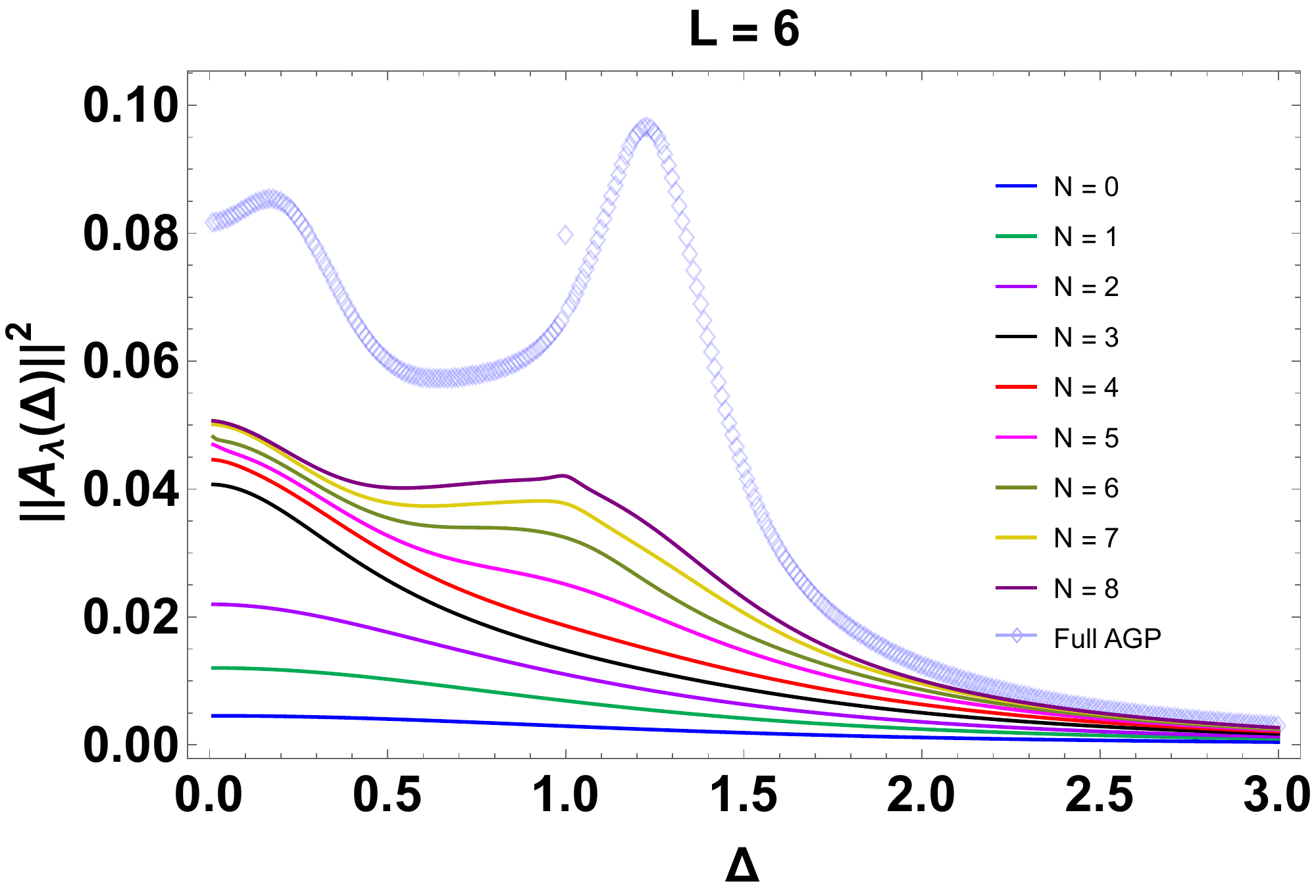}\label{fig:AGPXXZ06}}
	\hfill
	\subfigure[]{\includegraphics[height=5.7cm,width=1\linewidth]{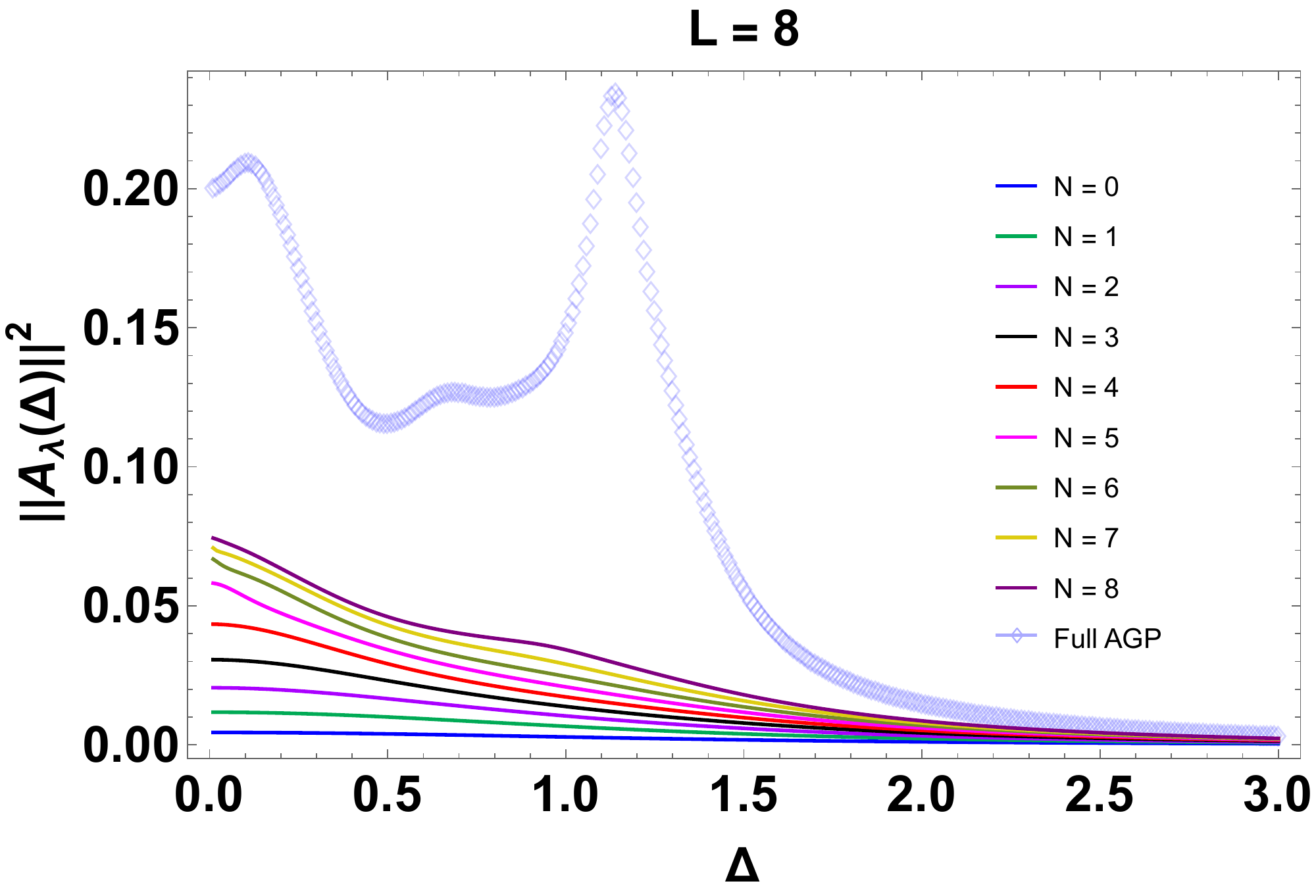}\label{fig:AGPXXZ08}}
	\hfill
	\subfigure[]{\includegraphics[height=5.7cm,width=1\linewidth]{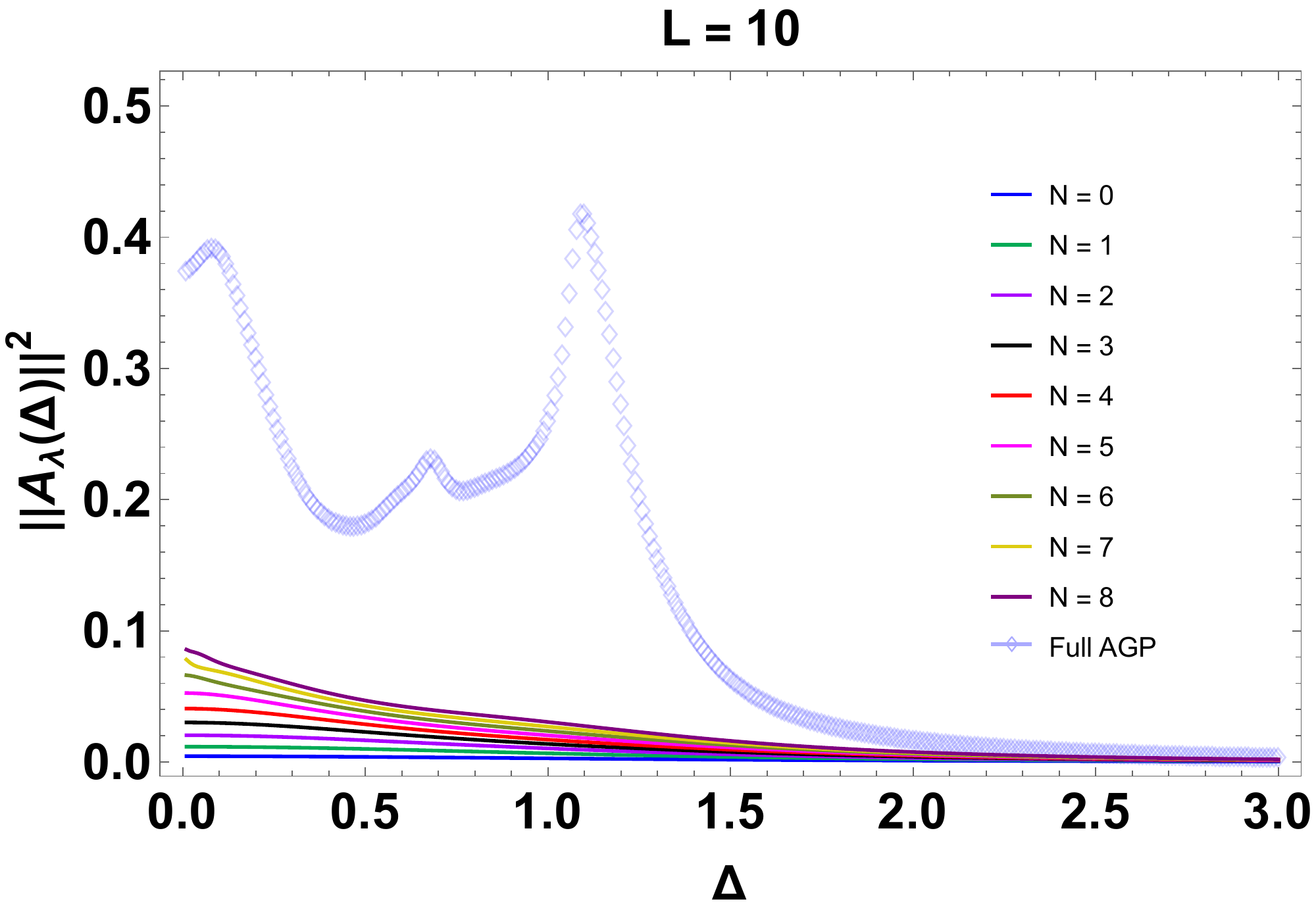}\label{fig:AGPXXZ10}}
	\caption{The AGP norm is compared by direct evaluation \eqref{agporg2} and by solving \eqref{mateqn} for different cutoffs and for system size (a) $L = 6$, (b) $L = 8$ and (c) $L = 10$. The cutoffs are taken at $N = 0,\dots, 8$. It is evident that taking such a small number of Lanczos coefficients fails to capture the AGP to a decent reasonable extent. For a small system size $L = 6$, the truncated AGP manages to barely replicate the behavior of the full AGP. However }\label{XXZAGP}
\end{figure}

We evaluate the AGP norm for this system via \eqref{agporg2} and compare it to the result obtained from \eqref{mateqn} by truncating successively at $N = 0$ to $N = 8$. We evaluate the AGP for $L = 6, 8$, and $10$. The numerical results are demonstrated in Fig.\,\ref{XXZAGP}

From Fig.\,\ref{fig:AGPXXZ06}, it can be seen that the $N = 8$ truncated AGP begins to show similar behavior to the full AGP. It is expected that the next few orders should be enough to replicate the qualitative nature of the AGP, though the full Krylov space will need to be spanned in order to exactly capture the AGP norm. The results become poorer with increasing $L$ (as seen in Fig.\,\ref{fig:AGPXXZ08} and Fig.\,\ref{fig:AGPXXZ10}) since the Krylov space dimensions increase while $N$ remains the same.

\subsection{Strongly Chaotic System}
To study the truncation method in a strongly chaotic system, we focus on the chaotic Ising model.
\begin{align}
	H = \sum_{i = 1}^{L}\sigma^{z}_{i}\sigma^{z}_{i+1} + h_{x}\sigma^{x}_{i} + h_{z}\sigma^{z}_{i}\label{HamIS}
\end{align}
with periodic boundary conditions. The model is known to be strongly chaotic for $h_{z} = \frac{\sqrt{5}+1}{4}$. The adiabatic deformation parameter is taken to be $h_{x}$.

\begin{figure}[htbp]
	\subfigure[]{\includegraphics[height=5.7cm,width=1\linewidth]{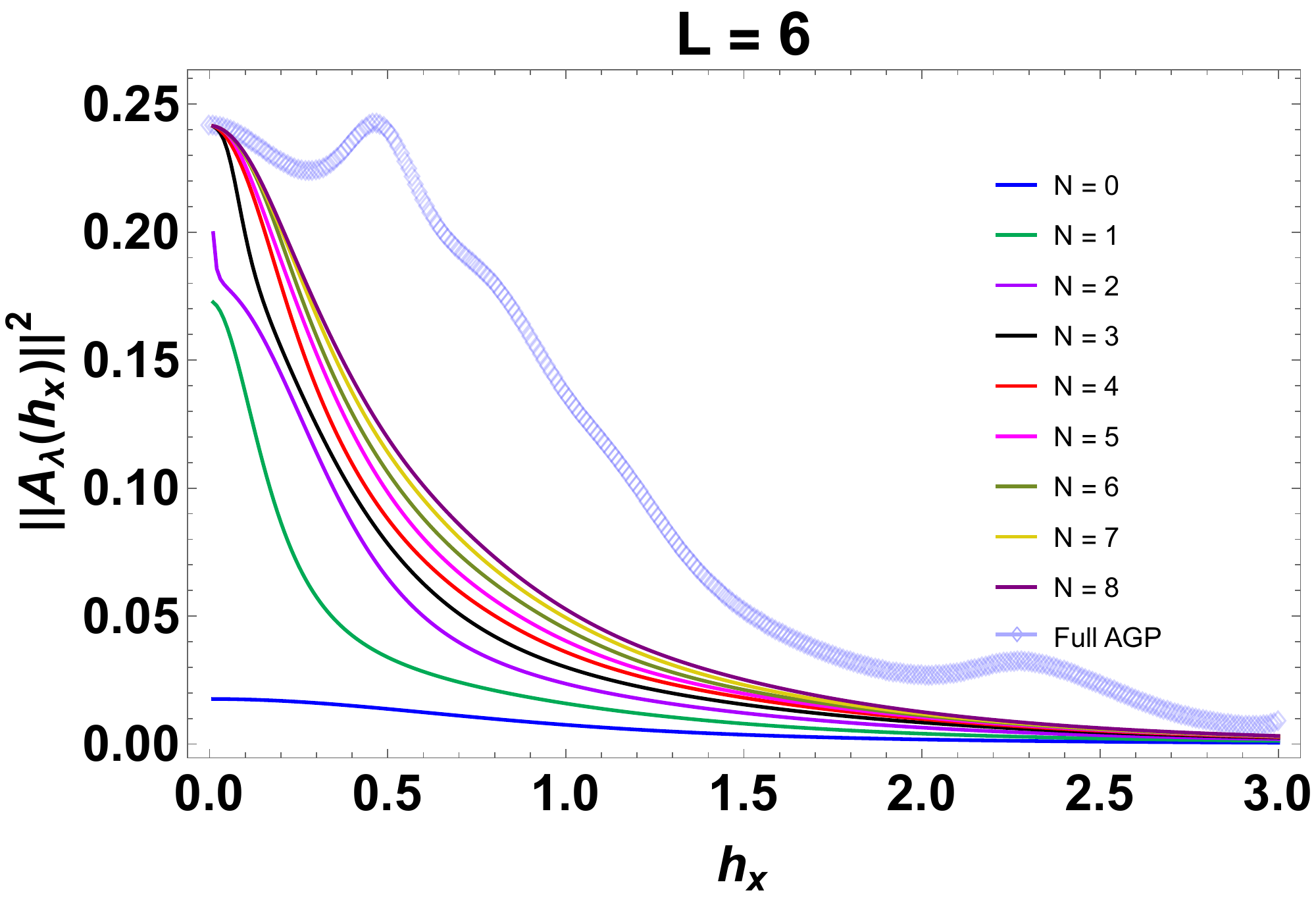}\label{fig:AGPChaIsing06}}
	\hfill
	\subfigure[]{\includegraphics[height=5.7cm,width=1\linewidth]{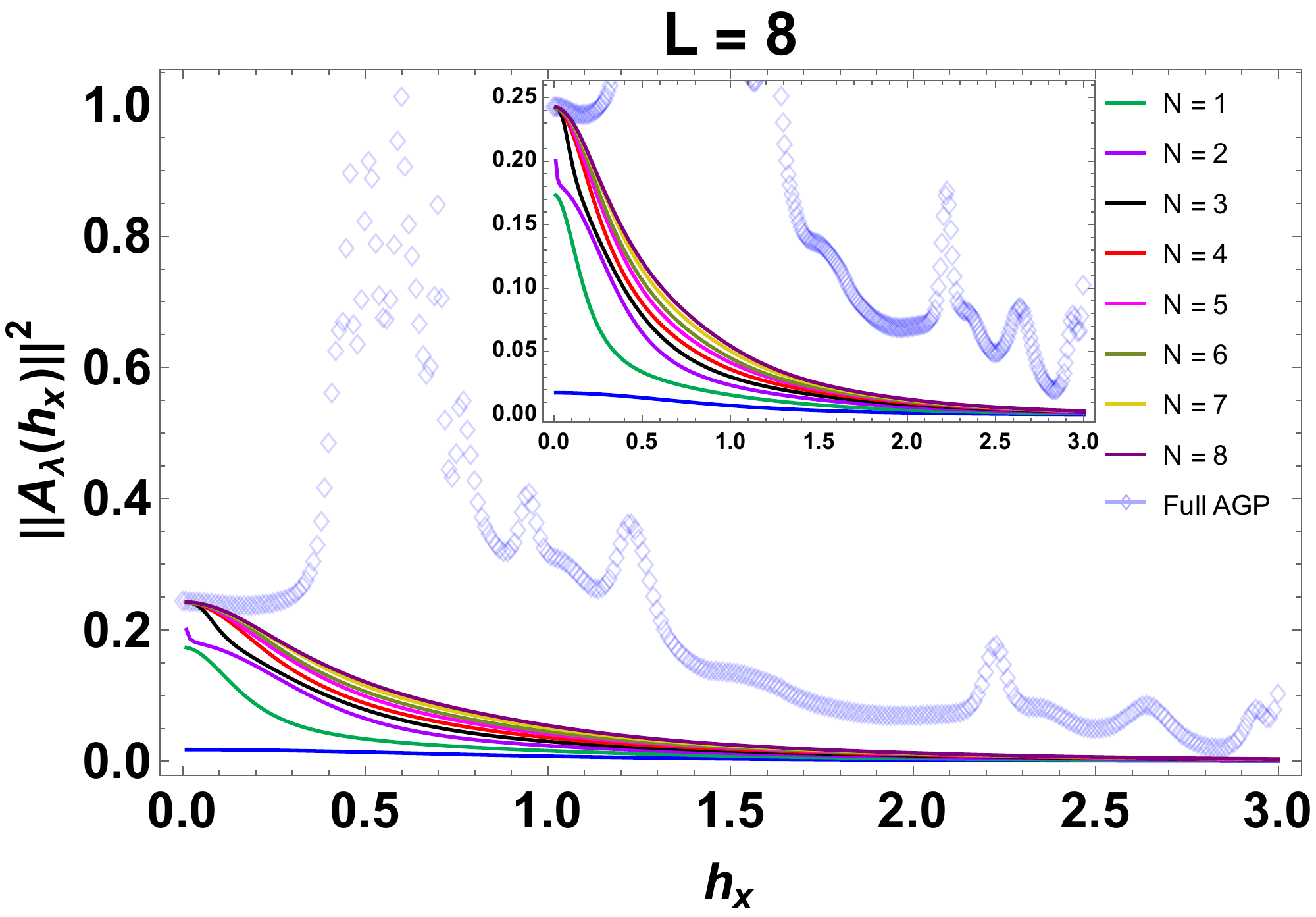}\label{fig:AGPChaIsing08}}
	\hfill
	\subfigure[]{\includegraphics[height=5.7cm,width=1\linewidth]{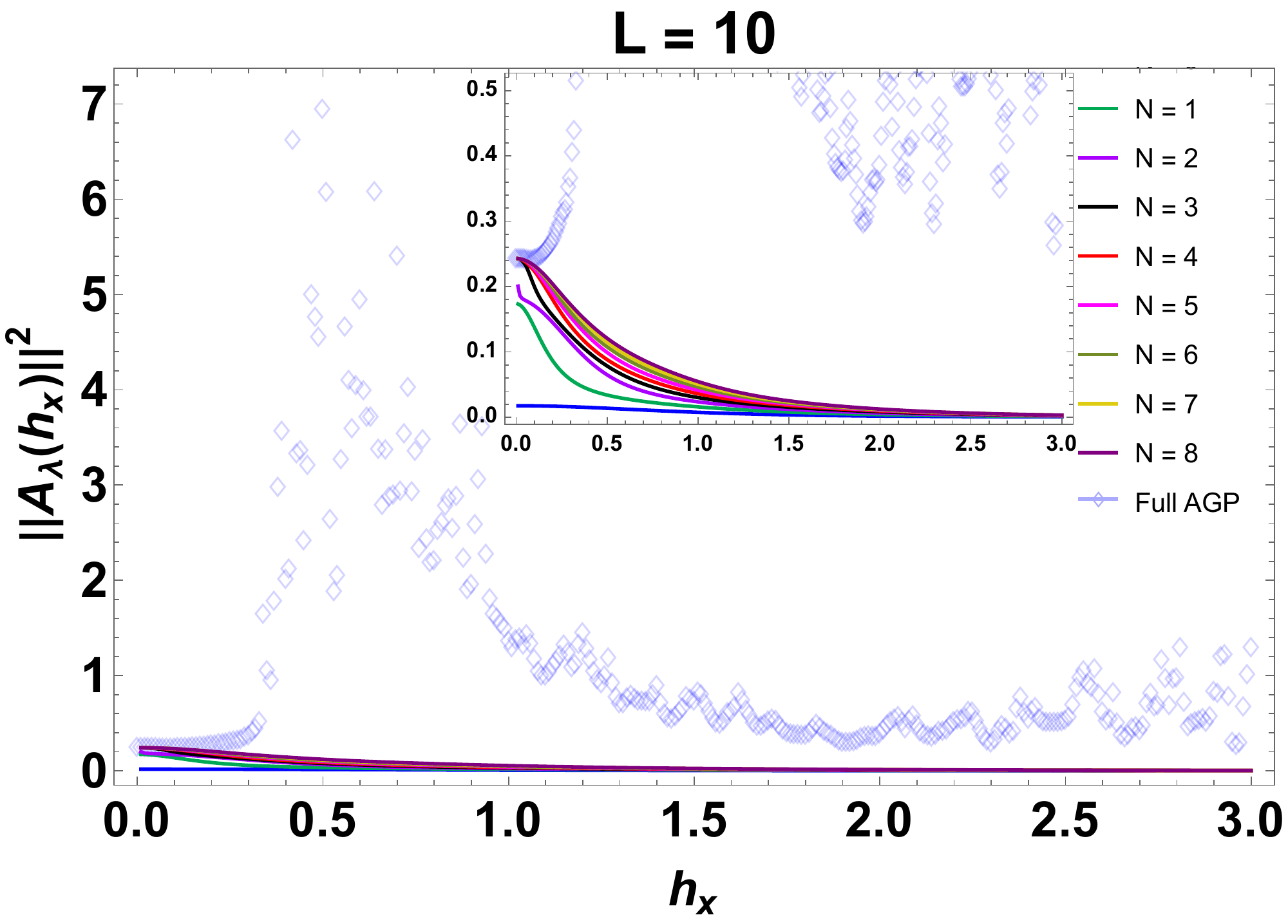}\label{fig:AGPChaIsing10}}
	\caption{The AGP norm is compared by direct evaluation \eqref{agporg2} and by solving \eqref{mateqn} for different cutoffs and for system size (a) $L = 6$, (b) $L = 8$ and (c) $L = 10$. The cutoffs are taken at $N = 0,\dots, 8$. It is evident that taking such a small number of Lanczos coefficients fails to capture the AGP to any reasonable extent. Inset in (b) and (c) shows a close-up of the results obtained via \eqref{mateqn}. }\label{ChaIsingAGP}
\end{figure}

The strongly chaotic nature of the system is also reflected by the fact that using \eqref{mateqn} up to $N = 8$ fails to capture the exact AGP. The Krylov space dimension for such systems are close to the bound $K \leq \mathcal{D}^2 - \mathcal{D} + 1$ \cite{Barbon:2019wsy,Rabinovici:2020ryf,Rabinovici:2021qqt}.  Therefore, a much larger set of Lanczos coefficients are required to be fed into \eqref{mateqn}, which implies that the knowledge of a much larger set of Krylov vectors is needed.

Making a naïve qualitative comparison, we find that the truncated AGP best captures the full AGP for the integrable (free) model \eqref{IntegrableIsingModel}, less so for the weakly chaotic LMG model \eqref{LMGHam}, even less so for the interacting $XXZ$ chain \eqref{HamXXZ} and least so for the chaotic Ising model \eqref{HamIS}. This is related to the fact that the size of the Krylov space increases in that order, and so the truncated AGP (which spans a part of the full Krylov space) captures the AGP to a reduced level of success.

\section{Probing Quantum Chaos: AGP and Operator Growth Hypothesis}
The norm of the regulated Adiabatic Gauge Potential has been shown to be sensitive to the degree of chaos in the system \cite{PhysRevX.10.041017}. The sensitivity of the AGP norm is the highest when the regulator is chosen to be exponentially suppressed by the system size. Specifically, for a system of size $L$, the optimal choice of the regulator is found to be $\mu = L 2^{-L}$. This choice also plays an important role in the scaling of the AGP norm with the system size. The rescaled AGP norm $\vert \vert A_\lambda \vert \vert^2 / L$ scales exponentially $e^{L}$ for chaotic systems, as $L^{\beta}$ (for some constant $\beta$) for integrable interacting systems and nearly constant (up to exponentially suppressed corrections) for integrable non-interacting systems. We shall study some toy models and check these statements by using \eqref{AGPC2}.

The first result we consider is the bound on the AGP norm. It was observed in \cite{PhysRevX.10.041017} that the AGP norm cannot grow faster than $4^M$ (for system size $M$). In other words, it is bounded above by a term proportional to $\frac{1}{\mu^2}$. It is simpler to consider \eqref{eq30}, and note that in the explicit form \eqref{eq31-2} one can use the fact that $\psi_{2 n + 1}(t) \leq 1 \;\; \forall \;n$
\begin{align}
	\vert \vert A_\lambda \vert \vert^2 \leq \vert \vert \partial_\lambda H \vert \vert^2 \sum_{n = 0}^{M}\int_{0}^{\infty}e^{-\mu (t+t')}\mathrm{d}t\mathrm{d}t' = \frac{M}{\mu^2}\vert \vert \partial_\lambda H \vert \vert^2
\end{align}
which confirms the bound. Here we have inserted the norm of the deforming operator $\vert \vert \partial_\lambda H \vert \vert^2$ by hand since \eqref{eq30} involves the normalized operator. Now, we begin by considering various types of autocorrelation functions and the corresponding AGP norms\footnote{We shall denote these by $A_\lambda (\mu)$ as before, although we will neglect any adiabatic parameter $\lambda$. So there will not be any $\lambda$ explicitly.}.

\subsubsection{Integrable: $b_{n} \sim \sqrt{n}$}
The autocorrelation function we study is $\mathcal{C}(t) = e^{-\frac{t^2}{2}}$. This corresponds to the Heisenberg-Weyl type Hamiltonian \cite{Caputa:2021sib}. An autocorrelation function of this form is found when considering the growth of the $\sigma_{z}$ operator under the integrable non-interacting Ising model Hamiltonian \eqref{IntegrableIsingModel} at criticality ($h  = 1$) \cite{Cao:2020zls}. The result we get is the following
\begin{align}
	\vert \vert A_{\lambda}(\mu) \vert \vert^2 = \frac{\sqrt{\frac{\pi }{2}} e^{\frac{\mu ^2}{2}} \left(\mu ^2+1\right) \text{Erfc}\left(\frac{\mu }{\sqrt{2}}\right)}{\mu }-1 \label{lab1}
\end{align}
where Erfc indicates the complementary error function. Taking an expansion around $\mu \rightarrow 0$ gives us the following asymptotic behavior
\begin{align}
	\frac{ \vert \vert A_\lambda(L)  \vert \vert^2}{K} \sim -\frac{2 \mu ^2}{3}+\frac{3}{4} \sqrt{\frac{\pi }{2}} \mu +\frac{\sqrt{\frac{\pi }{2}}}{2 \mu }-1\label{lab1asymp}
\end{align}
It can be seen that the function \eqref{lab1} (and consequently \eqref{lab1asymp}) grows as $e^{L}/L$ asymptotically. Considering the asymptotic growth of AGP with system size $L$ as the probe of chaotic dynamics \cite{PhysRevX.10.041017}, this result presents an apparent contradiction between the AGP and the Operator Growth Hypothesis results. This is also seen in the numerical result Fig.\,\ref{fig:AGPToy1}. The significance of this contradiction is not very clear at the moment. Part of the reason behind this may be that the AGP operator at $t = 0$ is usually a highly non-local operator. The autocorrelation function considered here has mostly been studied by considering the time evolution of local operators. Whether locality is indeed of significance or not (in this context) remains to be seen.  

There is also the issue of the norm of $\partial_\lambda H$. One may wonder if multiplying the result obtained from \eqref{AGPC2} by $\vert \vert \partial_\lambda H \vert \vert^2$ will change the scaling behavior. From most simple systems (such as the Ising model or the XXZ chain), it is straightforward to see that  $\vert \vert \partial_\lambda H \vert \vert^2 \sim O(L)$. This does not change our conclusions. Even if we allow the system to be such that $\vert \vert \partial_\lambda H \vert \vert^2 \sim O(L^{\beta})$ for some $\beta > 1$, the scaling with system size would still be exponential asymptotically.

\subsubsection{Chaotic: $b_{n} \sim n$}
\begin{figure}[htbp]
	\subfigure[]{\includegraphics[height=5.7cm,width=1\linewidth]{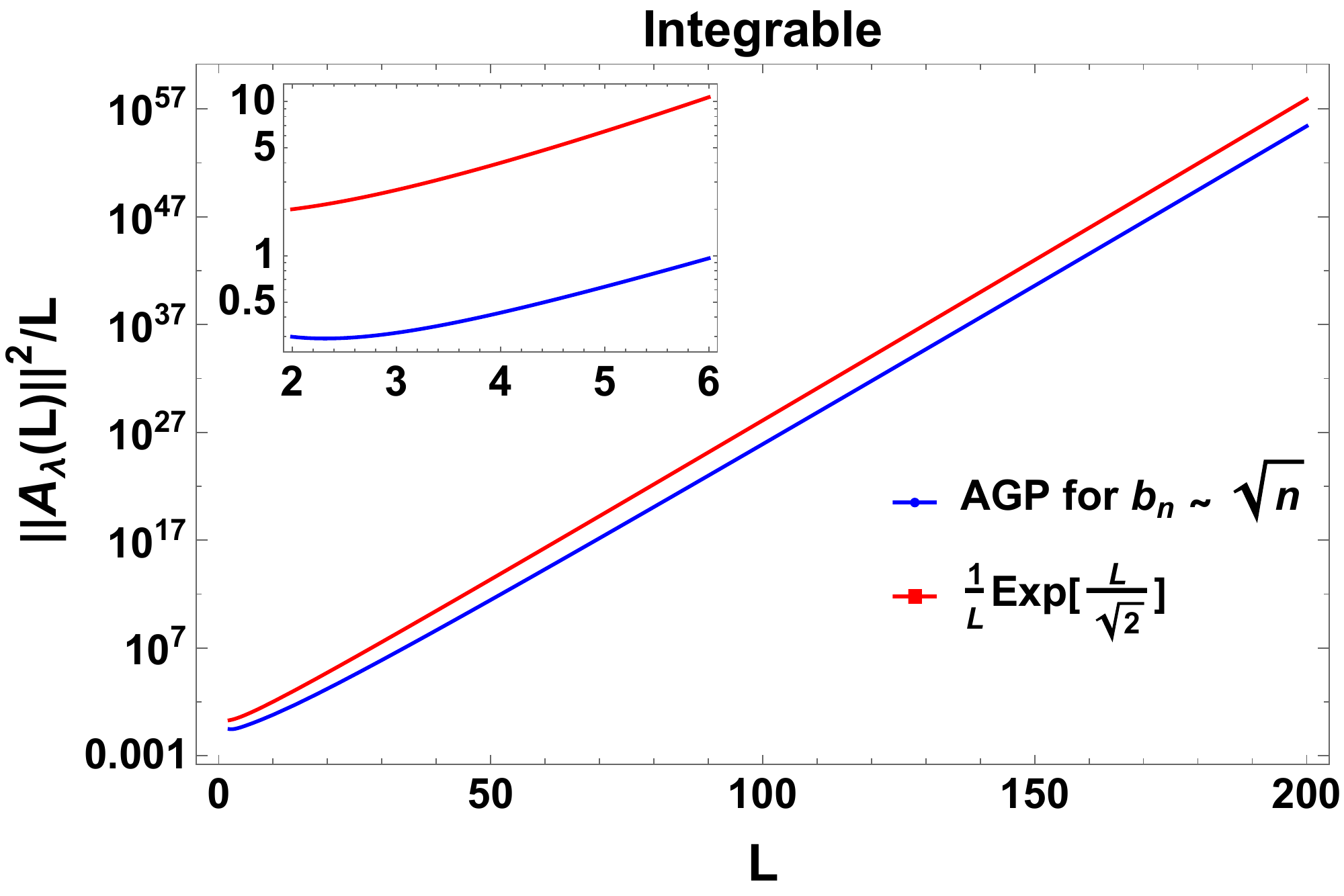}\label{fig:AGPToy1}}
	\hfill
	\subfigure[]{\includegraphics[height=5.7cm,width=1\linewidth]{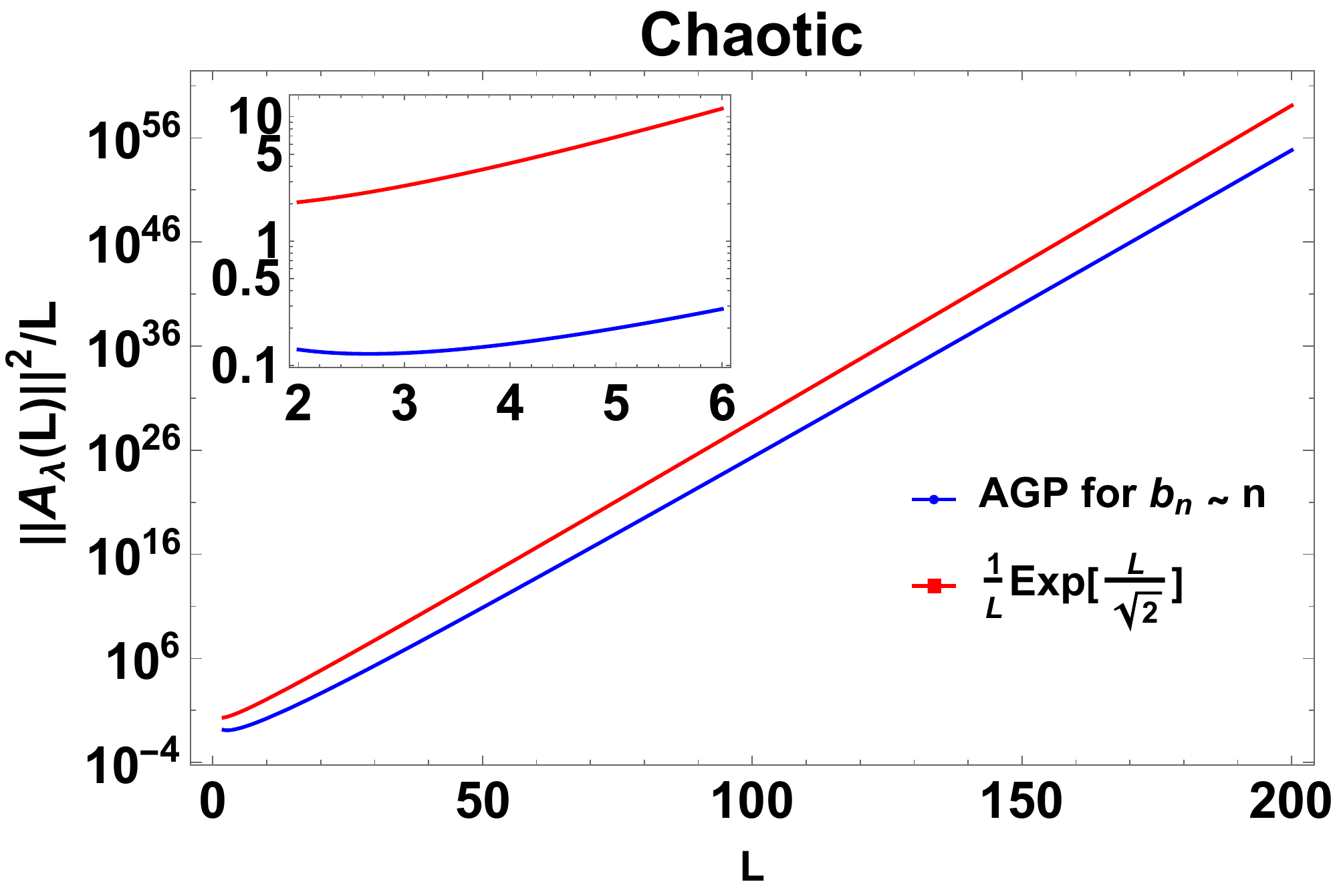}\label{fig:AGPToy2}}
	\caption{(a) The scaling of the AGP norm (Log plot) corresponding to a $\partial_\lambda H$ that gives rise to Lanczos coefficients that grow as $b_n \sim \sqrt{n}$, with system size $L$. According to the Operator Growth Hypothesis, this type of behavior (of the Lanczos coefficients) is associated with non-chaotic systems. (Inset): Scaling behavior for small $L$. (b) The scaling of the AGP norm (Log plot) for a $\partial_\lambda H$ corresponding to Lanczos coefficients that grow as $b_n \sim n$ asymptotically, with system size $L$. This type of behavior of the Lanczos coefficients is associated with chaotic systems, according to the Operator Growth Hypothesis. This plot corresponds to $\eta = 5$ and $\alpha = 4$. (Inset): Scaling behaviour for small $L$.} \label{AGPToy12}
\end{figure}

The autocorrelation function we study here corresponds to the asymptotically linear growth of Lanczos coefficients. The specific form of the Lanczos coefficients we choose are $b_{n} = \alpha\sqrt{n(n - 1 + \eta)}$ \cite{PhysRevX.9.041017}. The corresponding autocorrelation function has the form $\mathcal{C}(t) = \sech(\alpha t)^{\eta}$. The corresponding AGP norm comes out to be
\begin{widetext}
	\begin{align}
		\vert \vert A_\lambda(\mu) \vert \vert^2 = 2^{\eta -2} \Big(&\frac{\Gamma \left(\frac{1}{2} \left(\eta -\frac{\mu }{\alpha }\right)\right) \left(\Gamma \left(\frac{\alpha  \eta +\mu }{2 \alpha }\right)-\Gamma (\eta ) \, _2\tilde{F}_1\left(\eta ,\frac{1}{2} \left(\eta -\frac{\mu }{\alpha }\right);\frac{1}{2} \left(\eta -\frac{\mu }{\alpha }+2\right);-1\right)\right)}{\alpha  \mu  \Gamma (\eta )}\notag\\ &-\frac{2 \, _3F_2\left(\eta ,\frac{\eta }{2}+\frac{\mu }{2 \alpha },\frac{\eta }{2}+\frac{\mu }{2 \alpha };\frac{\eta }{2}+\frac{\mu }{2 \alpha }+1,\frac{\eta }{2}+\frac{\mu }{2 \alpha }+1;-1\right)}{(\alpha  \eta +\mu )^2}\Big)
	\end{align}
\end{widetext}
It is easy to see (by doing an expansion around $\mu = 0$) that the leading order term in $\mu$ is $\propto \frac{1}{\mu}$. Therefore, the system demonstrates chaotic scaling since $\mu \sim e^{-L \log 2}$. This is compatible with the operator growth hypothesis. The same is reflected in the numerical result Fig.\,\ref{fig:AGPToy2}.

\subsubsection{Finite Krylov space: $SU(2)$ symmetry}
Here we study a case where the Krylov space is highly restricted. This indicates strong integrability and has been found in real-life examples while studying the state complexity of many-body scars \cite{Bhattacharjee:2022qjw}. The Hamiltonian of such a system is given by
\begin{align}
	H = \alpha (J_{+} + J_{-})\label{HamSu2}
\end{align}
where $J_{+}$ and $J_{-}$ are the raising and lowering operators in the $SU(2)$ algebra.

\begin{figure}[htbp]
	\includegraphics[height=5.7cm,width=\linewidth]{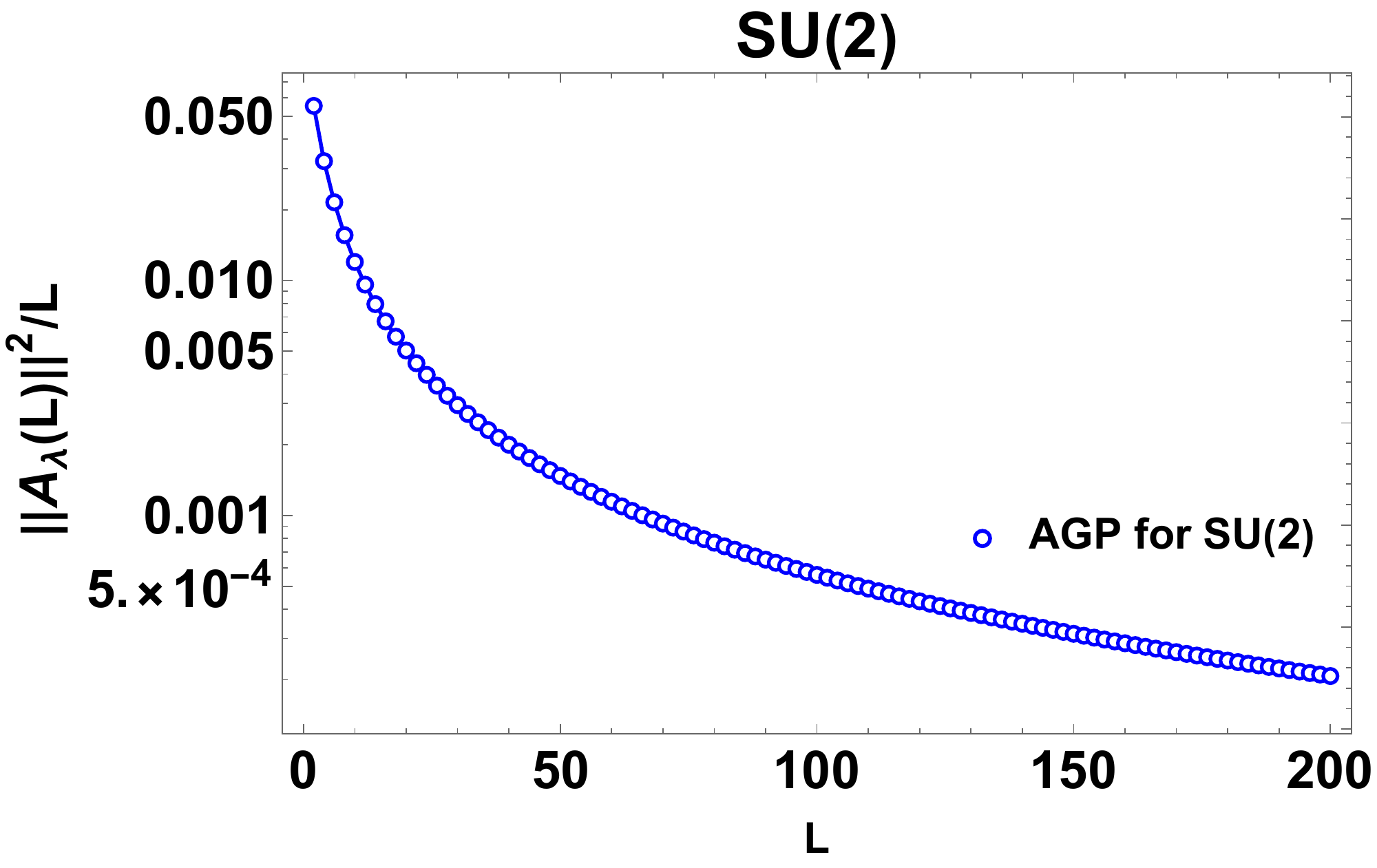}
	\caption{The AGP norm with the system size $L$ for the autocorrelation function corresponding to an $SU(2)$ Hamiltonian \eqref{HamSu2}. It is observed that the growth is not exponential, indicating integrable dynamics.}\label{fig:AGPToy3SU2}
\end{figure}

The autocorrelation function that we consider here (corresponding to the Hamiltonian described above) is given as $\mathcal{C}(t) = \cos^{L}\alpha t$ \cite{Caputa:2021sib}, where $L = 2 j$ (given the integer spin$-j$ representation of $SU(2)$). This autocorrelation function has the feature that the system size is a part of the autocorrelation function itself. The AGP norm turns out to have the following expression
\begin{align}
	&\vert \vert A_\lambda (L) \vert \vert^2 = \frac{2^{L-1}(\cosh (L \log (2))-\sinh (L \log (2)))}{L \left(2^{-L} L+i \alpha  L\right)^2}\times \notag\\ &\Big\{ 2^{-L} L \, _3F_2\left(-L,f_1(L),f_1(L);f_1(L)+1,f_1(L)+1;-1\right) \notag \\ &- \left(2^{-L} L+i \alpha  L\right) \, _2F_1\left(-L,f_1(L);f_1(L)+1;-1\right)\Big\}
\end{align}
where
\begin{align}
	f_1 (L) = \frac{i 2^{-L-1} L}{\alpha }-\frac{L}{2}
\end{align}
Since this is a complicated function, it is easier to simply plot the dependence of $A_\lambda (M)$ with $L$. Note that $L$ has to be even since $L = 2 j$ where $j$ is an integer spin. The result is given in Fig.\,\ref{fig:AGPToy3SU2}. The growth is sub-exponential, which reflects the highly restricted nature of the Krylov space. It is also a sign of the integrability of the system.

\subsubsection{Constant Lanczos: $b_{n} \sim \alpha$}
We consider a special kind of autocorrelation function \cite{noh2021operator}, which gives rise to a set of Lanczos coefficients that are constant. This autocorrelation function is $\mathcal{C}(t) = \frac{J_{1}(2\alpha t)}{\alpha t}$, where $J$ is the Bessel function of the first kind. It follows from this that the Lanczos coefficients are $b_{n} = \alpha$, which is a constant. This is an integrable system according to the Operator Growth Hypothesis. The AGP norm for this autocorrelation function becomes
\begin{align}
	\vert \vert A_\lambda \vert \vert^2 = \frac{\sqrt{\frac{4 \alpha ^2}{\mu ^2}+1}-1}{\alpha ^2}-\frac{2}{\mu  \sqrt{4 \alpha ^2+\mu ^2}}
\end{align}
A small $\mu$ expansion gives a series of the form $\vert \vert A_\lambda \vert \vert^2 = \frac{1}{\alpha  \mu }-\frac{1}{\alpha ^2} + O(\mu)$, which implies that the norm scales as $\frac{1}{\mu}$ and is thus identified as chaotic according to the AGP probe. The two probes do not agree.

\subsubsection{Oscillatory Lanczos}
A very interesting class of Lanczos coefficients are those that demonstrate different behavior for even and odd indices. In other words, $b_{2 k + 1} = f(k)$ and $b_{2k} = g(k)$, where $f$ and $g$ are different functions. There are some known autocorrelation functions that give rise to such Lanczos coefficients. One such autocorrelation function is $\mathcal{C}(t) = J_{0}(\alpha t)^2$. The AGP norm corresponding to this autocorrelation function is
\begin{align}
	\vert \vert A_\lambda \vert \vert^2 = -\frac{2 \left(E\left(-\frac{4 \alpha ^2}{\mu ^2}\right)-\left(\frac{4 \alpha ^2}{\mu ^2}+1\right) K\left(-\frac{4 \alpha ^2}{\mu ^2}\right)\right)}{\pi  \mu ^2 \left(\frac{4 \alpha ^2}{\mu ^2}+1\right)}
\end{align}
where $K$ is the complete elliptic integral of the first kind and $E$ is the incomplete elliptic integral of the second kind. 
The AGP norm goes as 
\begin{align}
	\vert \vert A_\lambda \vert \vert^2  = \frac{\log \left(\frac{64 \alpha ^2}{\mu ^2}\right)-2}{2 \pi  \alpha  \mu } + O(\mu)
\end{align} 
In terms of the system size $L$, the quantity $	\vert \vert A_\lambda \vert \vert^2/ L$ scales as $\frac{2^L}{L^2}$, which is a signature of chaotic behaviour. This is once more in contradiction with the result one expects from the Operator Growth Hypothesis.

\begin{figure}[htbp]
	\includegraphics[height=5.7cm,width=\linewidth]{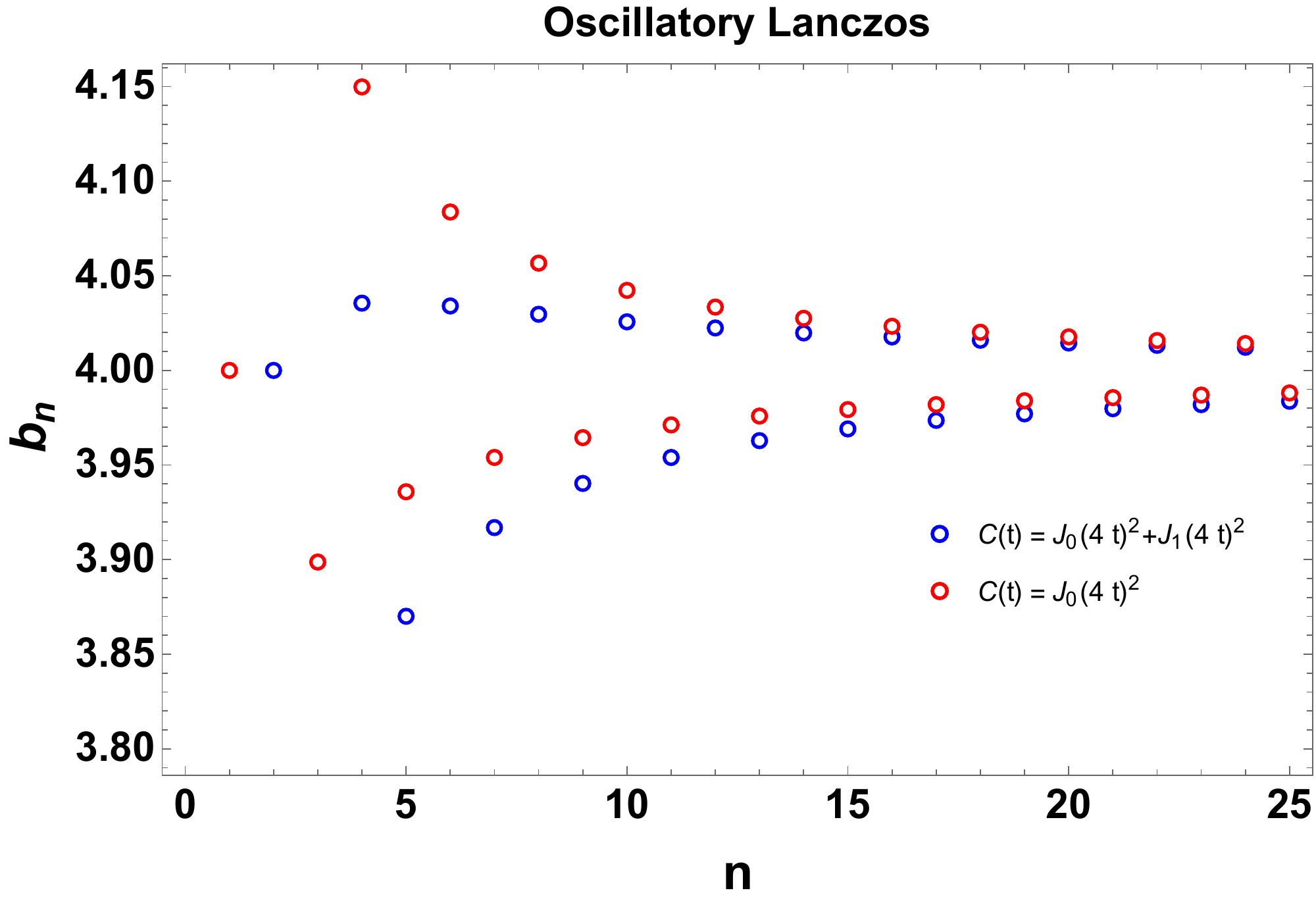}
	\caption{The Lanczos coefficients corresponding to the two autocorrelation functions $\mathcal{C}(t) = J_{0}(4 t)^2 + J_{1}(4 t)^2$ and $\mathcal{C}(t) = J_{0}(4 t)^2$ . The even and odd indexed coefficients show different behavior. These systems, according to the Operator Growth Hypothesis, are identified as integrable. The case is the opposite when using the AGP norm as the probe.}\label{fig:OscLanczos}
\end{figure}

Another closely related autocorrelation function is one that arises in the spin-$1/2$ $XY$ chain. The autocorrelation funcion $(\sigma^{x}_0 (t) \sigma^{x}_0)$ \cite{brandt1976exact} has the form
$\mathcal{C}(t) = J_{0}(4 t)^2 + J_{1}(4 t)^2$. This gives rise to another similar set of oscillating coefficients. The corresponding AGP norm takes the form.

\begin{align}
	\vert \vert A_\lambda \vert \vert^2 = \frac{\left(\mu ^2+32\right) K\left(-\frac{64}{\mu ^2}\right)-\mu ^2 E\left(-\frac{64}{\mu ^2}\right)}{8 \pi  \mu ^2}
\end{align}
Again, this scales as $\vert \vert A_\lambda \vert \vert^2 = \frac{-2 \log (\mu )-4+\log (1024)}{4 \pi  \mu } + O(\mu)$. This is once again indicative of chaotic dynamics according to the AGP probe. The two methods (Operator Growth Hypothesis and AGP probe) are in conflict. The oscillatory Lanczos coefficients for both these autocorrelation functions are shown in Fig \,.\ref{fig:OscLanczos}.

From the results derived in this section, we discover that in some toy models, the Operator Growth Hypothesis and the AGP probe (for chaos) are in apparent contradiction. This is a fairly naïve study, and a more systematic and exhaustive approach is required to reach a conclusion. One point to note here is that the AGP operator is a special type of operator in view of locality (and analogously, in terms of sparseness in its' matrix representation). The dependence of K-complexity on the type of operator choice is not very well understood yet. In that respect, the direct application of the Operator Growth Hypothesis that we undertake in this section may not be necessarily correct. Further investigation is required to reach a consensus. We defer it to future work.

\section{Conclusions}
The Adiabatic Gauge Potential has recently been considered as a probe of chaotic and integrable dynamics. Explicit expressions for the same, are hard to compute. It is only in certain integrable non-interacting systems that it has been possible to derive the expression. This difficulty arises partly because of the need to evaluate an infinite number of commutators. A way to bypass this process in some simple systems is by using a Gram-Schmidt-like orthonormalization procedure that brings down the number of computation steps to the bare minimum necessary \cite{hatomura2021controlling}. 

We have discussed a formalism based on the Lanczos algorithm, which is one such orthonormalization procedure, and which allows us to evaluate the AGP operator and the (regularized) AGP norm by studying the time evolution of the adiabatic deformation operator. We show that using the Lanczos algorithm is akin to choosing an optimal basis vis-à-vis the variational approach for evaluating the AGP \cite{selspol}. Using this, we introduce a matrix equation that can be solved to evaluate the (Regularised) AGP operator and norm. We discuss a truncated approach in which we only evaluate a few operators and study how well it matches the actual result. We find that the matching is very good for integrable non-interacting systems, reasonably good for weakly chaotic systems, and very poor for strongly chaotic systems. The reason behind that is the relative sizes of the Krylov space (generated out of the Lanczos algorithm) in the three situations, respectively.

The Lanczos algorithm, however, has another interesting use. It is the central ingredient in the formalism of Krylov complexity which is used as a probe of chaotic dynamics. We derive an expression for the AGP norm in terms of an integral transform of the autocorrelation function and a derivative of Green's function in the Krylov language. In the process, we find that the AGP response function provides a nice compatibility test between the Operator Growth Hypothesis and the Eigenstate Thermalization Hypothesis. We then evaluate the AGP for some systems with known autocorrelation functions. We compare the Operator Growth Hypothesis \cite{PhysRevX.9.041017} and the proposal for using AGP as a probe of chaos \cite{PhysRevX.10.041017}. We demonstrate that it is possible to construct some toy models where there is an apparent contradiction between the two probes. 

There are various open questions that are of interest and require investigation. A natural future direction is to apply the analog of the Lanczos algorithm for open quantum systems \cite{Bhattacharya:2022gbz, Bhattacharjee:2022lzy} to the AGP for open quantum systems \cite{Alipour_2020}. Regarding probing chaotic dynamics, the AGP and K-complexity are fundamentally different objects. The AGP probes the small-$\omega$ behavior (as seen from \eqref{AGP}) while the K-complexity probes the large-$\omega$ behavior as seen from the spectral function\footnote{We thank Xiangyu Cao for pointing this out.}. This is part of the reason why AGP is, at times, more effective than K-complexity at probing phase transitions. However, the AGP norm can still be formally related to the machinery of K-complexity. It would be interesting to attempt to understand one in terms of the other. It would also be interesting to understand the scaling behavior of the AGP norm for different types of systems (chaotic/integrable/interacting, etc.) from a more fundamental perspective. It would also be interesting to study in detail the compatibility issues between AGP and K-complexity with respect to probing chaos. It would be of interest to apply the Lanczos formalism to derive the AGP operators of other, more complicated systems. Finally, it is of conceptual interest to extend the notion of adiabatic deformations to field theories and holography via the AGP.

\textit{Note added :} Towards the finalizing stages of the manuscript, Ref.\cite{Takahashi:2023nkt} appeared on ArXiv, which proposes a nearly identical Lanczos formalism as ours.
\section{Acknowledgements}
The author would like to thank Xiangyu Cao, Jyotirmoy Mukherjee, Pratik Nandy, Tanay Pathak, Dario Rosa, and Samudra Sur for discussions on this and related projects, as well as for comments on this manuscript. The author is supported by the Ministry of Human Resource Development, Government of India, via the Prime Ministers' Research Fellowship.
\appendix
\section{Derivation of the Expressions for the AGP}
\label{appA}
In this Appendix, we describe the steps (that we skipped in the main text) regarding the derivation of the expression for the AGP norm.

In the discussion of the response function, we obtained the expression for the $\overline \vert f_\lambda (\omega) \vert^2$. We note that the AGP norm is written as \cite{PhysRevX.10.041017}
\begin{align}
    \vert \vert A_\lambda \vert \vert^{2} = \int_{-\infty}^{\infty}\mathrm{d}\omega \frac{\omega^{2}}{(\omega^2 + \mu^2)^2}\overline{\vert f_\lambda (\omega) \vert}^{2}\,.\label{AGPREP}
\end{align}
Using \eqref{AGPREP} we find the expression for the AGP. The integral over $\omega$ for the first term in \eqref{f40} is 
\begin{align}
    \int_{-\infty}^{\infty} \mathrm{d}\omega \frac{\omega^2}{(\omega^2 + \mu^2)^2}e^{i\omega t} = \frac{\pi  e^{-\mu  | t| }}{2 \mu }-\frac{1}{2} \pi  | t|  e^{-\mu  | t| }
\end{align}
The second term in \eqref{f40} vanishes due to the $\delta(\omega)$ (note that $\mu > 0$). Therefore the expression for the AGP norm becomes
\begin{align}
    \vert \vert A_\lambda \vert \vert^{2} = \frac{1}{2}\int_{0}^{\infty} \mathrm{d}t \left(\frac{1}{\mu} - t\right)\mathcal{C}(t)e^{-\mu t}\,.\label{AGPCSome}
\end{align}
where we have changed the limits on the $t$ integral from $\{-\infty, \infty\}$ to $\{0, \infty\}$. Without the $\frac{1}{\mu}$ piece in the integrand, the AGP norm is simply the negative of the Laplace transform of the function $t\mathcal{C}(t)$. . The expression \eqref{AGPCSome} can also be obtained from \eqref{AGPC} by an appropriate change of variables.

\subsection{An alternative expression}
We demonstrate that there is an alternative expression in terms of the difference between two shifted autocorrelation functions, which we derive below.

We begin with the following expression
\begin{align}
    \vert \vert A_{\lambda} \vert \vert ^{2} &= \frac{1}{\mathcal{D}}\sum_{m \neq n}\vert\bra{m}A_{\lambda}\ket{n}\vert^{2} \notag \\ &= 
    \frac{1}{\mathcal{D}}\sum_{m \neq n}\frac{\omega^{2}_{m n}}{(\mu^{2} + \omega^{2}_{m n})^{2} }\vert \bra{m}\partial_{\lambda} H \ket{n}\vert^{2}\,. \label{agporg}
\end{align}
where $\mathcal{D}$ is the dimension of the Hilbert space. It is straightforward to show that this expression can be obtained from \eqref{AGP} directly. 

From the expression \eqref{AGPC1}, we now evaluate $\vert \vert A_\lambda \vert \vert^{2}$. The expression we obtain is 
\begin{align}
    \bra{m}A_\lambda\ket{k} = -\sum_{n = 0}^{M}\alpha_{2 n + 1}\bra{m}\mathcal{O}_{2 n + 1}\ket{k}\,. \label{exA}
\end{align}
Note that $\alpha_{2 n + 1}$ is the Laplace transform of $i^{2n + 1}\psi_{2 n + 1}$. It is simple to note that $\alpha^{\ast}_{2 n + 1} = - \alpha_{2 n + 1}$ by using the fact that $\psi_{2 n + 1}(t)$ are real functions of $t$. Noting that all the odd-indexed Krylov basis operators are anti-hermitian, the complex conjugate of \eqref{exA} is given by
\begin{align}
    \bra{m}A_\lambda\ket{k}^\ast &= -\sum_{n = 0}^{M}\alpha^\ast_{2 n + 1}\bra{k}\mathcal{O}_{2 n + 1}^\dagger\ket{m} \notag \\
    &= -\sum_{n = 0}^{M}\alpha_{2 n + 1}\bra{k}\mathcal{O}_{2 n + 1}\ket{m}\mathrm{d}t
\end{align}
Therefore, one can now write an expression for the regularized AGP matrix element as
\begin{align}
    \vert \bra{m}A_\lambda\ket{k} \vert^{2} = \sum_{n, n' = 0}^{M} &\alpha_{2 n + 1}\alpha_{2 n' + 1}\bra{k}\mathcal{O}_{2 n + 1}\ket{m}\notag \\ &\times\bra{m}\mathcal{O}_{2 n' + 1}\ket{k}
\end{align}
Finally, we have to sum over the eigenstates $\ket{m}, \ket{k}$ while keeping $m \neq k$. This leads us to the following expression
\begin{align}
    \sum_{m \neq k} \vert \bra{m}A_\lambda\ket{k} \vert^{2} &= \sum_{k}\sum_{n, n' = 0}^{M} \alpha_{2 n + 1}\alpha_{2 n' + 1}\notag \\ &\times \bra{k}\mathcal{O}_{2 n + 1}\left(\mathbf{I} - \ket{k}\bra{k}\right)\mathcal{O}_{2 n' + 1}\ket{k} \notag \\
    &= \sum_{k}\sum_{n, n' = 0}^{M} \alpha_{2 n + 1}\alpha_{2 n' + 1}
    \notag \\ &\times \bra{k}\mathcal{O}_{2 n + 1}\mathcal{O}_{2 n' + 1}\ket{k}
\end{align}
where in the last step, we have used $\bra{k}\mathcal{O}_{2 n + 1}\ket{k} = 0$. 

Lastly, we use the fact that 
\begin{equation}
\hspace{-1em}\sum_{k} \bra{k} \mathcal{O}_{2 n + 1}\mathcal{O}_{2 n' + 1} \ket{k} = -\text{Tr}\{\mathbf{1}\}(\mathcal{O}_{2 n + 1}\vert \mathcal{O}_{2 n' + 1})
 \end{equation}

with $\tr{\mathbf{1}} = \mathcal{D}$ and that all the Krylov basis operators $\mathcal{O}_{n}$ are orthonormal to each other. This gives us
\begin{align}
    \vert \vert A_\lambda \vert \vert^{2} = -\sum_{n = 0}^{M}\alpha_{2 n + 1}^{2}\label{eq30-2}
\end{align}
From the expression \eqref{al}, we find that this expression may be written as
\begin{align}
    \vert \vert A_\lambda \vert \vert^{2} = \sum_{n = 0}^{M}\int_{0}^{\infty}\psi_{2 n + 1}(t)\psi_{2 n + 1}(t')e^{-\mu(t + t')}\mathrm{d}t \mathrm{d}t'\label{eq31-2}
\end{align}
In order to arrive at the alternative expression, it is helpful to define the following term
\begin{align}
    \beta_{n} = \frac{1}{2}\int_{t = -\infty}^{\infty}\sg{t}i^{n}\left\{ \begin{aligned}
    &1 \;\; n \in \text{odd} \\
    &i \;\; n \in \text{even}
    \end{aligned} \right\}e^{-\mu |t|}\psi_{n}(t)\mathrm{d}t
\end{align}
Note that since $\beta_{2 n} = 0$ and $\beta_{2 n + 1} = \alpha_{2 n + 1}$, we can rewrite the AGP norm as
\begin{align}
    \vert \vert A_\lambda \vert \vert^2 = -\sum_{n = 0}^{K}\beta^{2}_{n}
\end{align}
This expression, when expanded out, becomes
\begin{align}
    \vert \vert A_\lambda \vert \vert^2 = \frac{1}{4}\sum_{n = 0}^{K}\int_{-\infty}^{\infty}\sg{t}\sg{t'}\psi_{n}(t)\psi_{n}(t')e^{-\mu |t| - \mu |t'|}\mathrm{d}t \mathrm{d}t'\,.\label{app0}
\end{align}
Note the sum $\sum_{n = 0}^{K}\psi_{n}(t)\psi_{n}(t')$. This sum equals $\psi_{0}(t - t')$ by definition, which is the autocorrelation function $\mathcal{C}(t-t')$. Intuitively, it follows from the fact that the sum of the product of two eigen-solutions of a differential equation is Green's function. This allows us to write the AGP norm as
\begin{align}
    \vert \vert A_\lambda \vert \vert^2 = \frac{1}{4}\int_{-\infty}^{\infty}\sg{t}\sg{t'}\mathcal{C}(t-t')e^{-\mu (|t| + |t'|)}\mathrm{d}t \mathrm{d}t'\label{app1}
\end{align}
It is worth simplifying this expression further. To do that, we simply turn the integrals over $t$ and $t'$ to the limits $\{0, \infty\}$. This gives us
\begin{align}
\vert \vert A_\lambda \vert \vert^2 = \frac{1}{2}\int_{0}^{\infty}\left(\mathcal{C}(t - t') - \mathcal{C}(t + t')\right)e^{-\mu ( t + t')}\mathrm{d}t\mathrm{d}t\,. \label{AGPC}
\end{align}
where we have used the fact that $\mathcal{C}$ is an even function.

\subsection{AGP as a product of Lanczos coefficients}
The moments of the autocorrelation function and the Lanczos coefficients contain the same information, repackaged in a different way \cite{PhysRevX.9.041017}. An iterative algorithm \cite{viswanath1994recursion} can be used to obtain one from the other. Due to that, it is instructive to consider the expression \eqref{AGPC2} and use the expansion of autocorrelation function in terms of moments \cite{PhysRevX.9.041017}. This gives us the following result
\begin{align}
    \vert \vert A_\lambda \vert \vert^2 = \sum_{n}\frac{n(-1)^{n+1} m_{2 n}}{\mathbf{\mu}^{2n + 2}}
\end{align}
Therefore, the knowledge of the moments of the autocorrelation function is equivalent to the knowledge of the AGP norm.

It is also possible to evaluate it in terms of the Lanczos coefficients themselves. In order to achieve that, we note the following relation between the moments $m_{2n}$ and the Lanczos coefficients $b_{n}$. The relation between moments and Lanczos coefficients is obtained via a product over Dyck paths. Using the expression in \cite{PhysRevX.9.041017}, we write the AGP norm as
\begin{align}
    \vert \vert A_\lambda \vert \vert^2 = \sum_{n}\frac{n (-1)^{n + 1}}{\mu^{2 n + 2}}\sum_{\{ c_{k}\} \in \mathcal{F}_{n}}\prod^{2 n}_{k = 1}b_{(c_k + c_{k - 1})/2}
\end{align}
The indices $c_{k} \geq \frac{1}{2}$ and $c_{0} = c_{2 n} = \frac{1}{2}$, with the constraint that $\vert c_{k} - c_{k + 1} \vert = 1$. The set of such paths is denoted by $\mathcal{F}_{n}$. The number of Dyck paths for a fixed $n$ is given by the Catalan number $C_n = \frac{\Gamma(2n+1)}{\Gamma(n+1)\Gamma(n+2)}$. As mentioned in \cite{PhysRevX.9.041017}, this provides a set of bounds on the moments $m_{2n}$, which is given by $b_{1}^2 b_{2}^2 \dots b_{n}^2\leq m_{2 n}\leq \text{max}_{k = 1}^{n} (b_{k}^{2}) C_{n}$. 

\section{Derivation of AGP operator for the analytic cases}
\label{appB}
In this Appendix, we detail the derivation of the AGP operator for the cases considered in the main text. We begin with considering the simple $2$-level system \eqref{2levsys}.
\subsection*{1. The $2$-level system}
\label{appB2l}
We employ the Lanczos algorithm to determine the AGP operator \eqref{AGPC1}. The steps are outlined below
\begin{itemize}
    \item Consider the Hamiltonian $H = \lambda \sigma^z + \Delta \sigma^x$ and the intial operator $\mathcal{O}_0 = \sigma^z \equiv \partial_\lambda H$. Note that this operator is normalized. 
    \item The first Krylov basis operator is obtained as $b_1 \mathcal{O}_1 = [H, \mathcal{O}_0]$. This gives us $\mathcal{O}_1 = - i \sigma^y$ and $b_1 = 2 \Delta$.  
    \item The second Krylov basis operator is written as $b_2 \mathcal{O}_2 = [H, \mathcal{O}_1] - b_1\mathcal{O}_0$. This gives us $b_2 = 2 \lambda$ and $\mathcal{O}_2 = - \sigma^x$. 
    \item The third Krylov basis operator turns out to be $b_3 \mathcal{O}_3 = [H, \mathcal{O}_2] - b_2 \mathcal{O}_1 = 0$. Therefore, we find that $b_3 = 0$, and the Lanczos algorithm terminates.
\end{itemize}
Therefore, the Krylov basis has three independent basis elements, denoted by $\mathcal{O}_0, \mathcal{O}_1$, and $\mathcal{O}_2$. The dimensions of the Krylov space is $3$, which saturates the bound $K \leq \mathcal{D}^2 - \mathcal{D} + 1$, where $K$ is the dimension of the Krylov space and $\mathcal{D}$ is the dimension of the Hilbert space. In this case $\mathcal{D} = 2$ and $K = 3$. 

Now that we have obtained the Krylov basis vectors, we turn out attention to the wavefunctions. These are obtained via the recursive relation \eqref{sch}. Note that \eqref{AGPC1} tells us that only the \textit{odd indexed} wavefunctions contribute to the AGP operator. Therefore, we only need to consider $\psi_1$, since $\psi_0, \psi_2$ do not contribute and all other $\psi_n\, \forall \,n\geq 3$ are $0$. From \eqref{sch}, it can be easily seen that 
\begin{align}
    \psi_1 = - \frac{1}{b_1}\partial_t \psi_0 (t)
\end{align}
Thus the AGP operator is written as (using \eqref{AGPC1})
\begin{align}
    A_\lambda &=  -\left(\int_{0}^{\infty}i e^{-\mu t} \psi_{ 1}(t)\mathrm{d}t\right)\mathcal{O}_{1} \notag \\
    &= \frac{1}{b_1}\left(\int_{0}^{\infty}i e^{-\mu t} \partial_t \psi_{ 0}(t)\mathrm{d}t\right)\mathcal{O}_{1}
\end{align}
Knowing the autocorrelation function $\psi_0 (t)$ allows us to evaluate the integral above. Integrating by parts turns this integral into an integral over $\psi_0 (t)$ only, which can be evaluated easily. The result turns out to be
\begin{align}
    A_\lambda &= -\frac{i}{b_1}(-i \sigma^y) \frac{4 \Delta^2}{4 \Delta^2 + 4 \lambda^2} \notag \\
    &= -\frac{1}{2} \frac{\Delta}{\Delta^2 + \lambda^2}\sigma^y
\end{align}
This result matches with the expression derived in \cite{PhysRevLett.109.115703}, as expected.  
\subsection*{2. A 2-qubit system}
\label{appB2q}
In this section, we derive the AGP operator for the Hamiltonian \eqref{2q}
\begin{align}
    H(\lambda) = - \left(\sigma^{1}_{x}\sigma^{2}_{x} + \sigma^{1}_{z}\sigma^{2}_{z} \right) - \varepsilon (1 - \lambda)\left( \sigma^{1}_{z} + \sigma^{2}_{z} \right)\,.\label{2q2}
\end{align}
The steps are the same as discussed for the $2-$ level system, so we shall omit them here. The results for the Krylov basis operators are as follows 
\begin{align}
    \mathcal{O}_{0} &= \frac{\partial_\lambda H}{\sqrt{\ipr{\partial_\lambda H}{\partial_\lambda H}}} = \frac{1}{\sqrt{2}}(\sigma^{1}_{z} + \sigma^{2}_{z}) \\
    b_{1}\mathcal{O}_{1} &= \frac{1}{b_1}\left([H,\mathcal{O}_0]\right) = \frac{i}{\sqrt{2}}(\sigma^{1}_{x}\sigma^{2}_{y} + \sigma^{1}_{y}\sigma^{2}_{x}) \\
    \mathcal{O}_{2} &= \frac{1}{b_2}\left([H,\mathcal{O}_1] - b_{1}\mathcal{O}_0\right) = \frac{1}{\sqrt{2}}\left(\sigma^{1}_{y}\sigma^{2}_{y} - \sigma^{1}_{x}\sigma^{2}_{x} \right)\\
    b_{3}\mathcal{O}_{3} &= [H,\mathcal{O}_2] - b_{2}\mathcal{O}_1 = 0
\end{align}
The Lanczos coefficients are $b_{1} = 2,\, b_{2} = 4 \vert \varepsilon ( 1 - \lambda) \vert$ and  $b_{n} = 0 \,\, \forall n \geq 3$. Therefore, the AGP operator is proportional to $\mathcal{O}_1$. Using the expression for the autocorrelation function \eqref{auc2q}, we have the following expression for $\alpha_1$
\begin{align}
    -i\int_{0}^{\infty}e^{-\mu t}\psi_{1}(t)\mathrm{d}t &= \frac{i}{b_{1}}\int_{0}^{\infty}e^{-\mu t}\partial_{t}\psi_{0}(t)\mathrm{d}t \notag \\
    &= - \frac{2 i}{4 + 16\varepsilon^2 (1 -\lambda)^2 + \mu^2}
\end{align}
It is possible to set $\mu \rightarrow 0$ in this expression, as the Hamiltonian \eqref{2q} is non-degenerate. This gives us
\begin{align}
    A_\lambda = -\frac{i \sqrt{2}}{4(1 + 4\varepsilon^2 ( 1 - \lambda)^2)}\left(\sigma^{1}_{x}\sigma^{2}_{y} + \sigma^{1}_{y}\sigma^{2}_{x}\right)
\end{align}
This result does not match exactly with \cite{petiziol} since we started with a \textit{normalized} $\partial_{\lambda}H$. To obtain the result derived in \cite{petiziol}, we must multiply this AGP operator by the norm of $\partial_\lambda H = \sqrt{2}\varepsilon$. This gives us the AGP operator
\begin{align}
    A_\lambda = -\frac{i \varepsilon}{2(1 + 4\varepsilon^2 ( 1 - \lambda)^2)}\left(\sigma^{1}_{x}\sigma^{2}_{y} + \sigma^{1}_{y}\sigma^{2}_{x}\right)
\end{align}
which is the result derived in \cite{petiziol}.
\subsection*{3. A 4-body system}
\label{appB4}
We consider the Hamiltonian \eqref{4body} and employ the Lanczos algorithm. Our starting point is the intial operator (normalized) $\partial_\lambda H(\lambda) = \frac{1}{\sqrt{2}}(\sigma^{z}_1 + \sigma^{z}_2)$. The Krylov basis vectors are evaluated in the same way as before, and their expressions are listed below
\begin{widetext}
\begin{align}
    \mathcal{O}_0 &= \frac{1}{\sqrt{2}}\left(\sigma^{z}_1 + \sigma^{z}_2 \right) \\
    \mathcal{O}_{1} &= -\frac{i}{2}\left(\sigma^{x}_0 \sigma^{y}_1 + \sigma^{x}_1 \sigma^{y}_2 + \sigma^{y}_1 \sigma^{x}_2 + \sigma^{y}_2 \sigma^{x}_3 \right) \\
    \mathcal{O}_{2} &= \frac{1}{\sqrt{8 + 10 \lambda^2}}\left(2\sigma^{x}_0 \sigma^{z}_{1}\sigma^{x}_2 + 2\sigma^{x}_1\sigma^{z}_2\sigma^{x}_3 + \lambda\left(2 \sigma^{y}_1\sigma^{y}_2 - 2\sigma^{x}_1\sigma^{x}_2 - \sigma^{x}_0 \sigma^{x}_1 - \sigma^{x}_2 \sigma^{x}_3 \right) \right) \\
    \mathcal{O}_{3} &= \frac{ 6 \lambda^2}{b_{2} b_{3}}\left(\mathcal{O}_{1} + i(\sigma^{x}_0 \sigma^{y}_1 + \sigma^{y}_2 \sigma^{x}_{3}) \right) + \frac{8 i \lambda}{b_2 b_3}\left(\sigma^{x}_0 \sigma^{z}_{1}\sigma^{y}_{2} + \sigma^{y}_{1}\sigma^{z}_{2}\sigma^{x}_{3} \right) \\
    \mathcal{O}_4 &= \frac{6 \lambda^2 b_2}{b_2 b_3 b_4} \mathcal{O}_2 + \frac{(12 \lambda^3 + \lambda b_3^2)}{b_2 b_3 b_4}\left( \sigma^{x}_0 \sigma^{x}_1 + \sigma^{x}_2 \sigma^{x}_3 \right) + \frac{(32 \lambda - 2 \lambda b_3^2)}{b_2 b_3 b_4}\sigma^{y}_1 \sigma^{y}_2 + \frac{2 \lambda b_3}{b_2 b_4}\sigma^{x}_1 \sigma^{x}_2 \notag \\ &+ \frac{(4 \lambda^2 - 2 b_3^2)}{b_2 b_3 b_4}\left(\sigma^{x}_0 \sigma^{z}_1 \sigma^{x}_2 + \sigma^{x}_1 \sigma^{z}_2 \sigma^{x}_3 \right) - \frac{ 32 \lambda}{b_2 b_3 b_4}\sigma^{x}_0 \sigma^{z}_1 \sigma^{z}_2 \sigma^{x}_3 \\
    \mathcal{O}_5 &= \left(\frac{6 \lambda^2 - b_4^2}{b_4 b_5} \right)\mathcal{O}_3 + \left(\frac{6\lambda^2 b_2^2 + 16\lambda^2 - 8 b_3^2 - 4 \lambda^2 b_3^2}{b_2 b_3 b_4 b_5}\right)\mathcal{O}_1 + \frac{24 i \lambda^4}{b_2 b_3 b_4 b_5}\left(\sigma^{x}_0 \sigma^{y}_1 + \sigma^{y}_2 \sigma^{x}_3 \right) \notag \\ &+ \left(\frac{6 i \lambda^2 b_3^2 - 64 i \lambda^2}{b_2 b_3 b_4 b_5} \right)\left(\sigma^{x}_1\sigma^{y}_2 + \sigma^{y}_1 \sigma^{x}_2 \right) + \left(\frac{128 i\lambda - 8 i \lambda b_3^2 + 8 i \lambda^3}{b_2 b_3 b_4 b_5} \right)\left(\sigma^{x}_0 \sigma^{z}_1 \sigma^{y}_2 + \sigma^{y}_1 \sigma^{z}_2 \sigma^{x}_3 \right) \\ 
    \mathcal{O}_6 &= \left(\frac{6 \lambda^2 - b_4^2 - b_5^2}{b_5 b_6} \right)\mathcal{O}_4 + \left(\frac{2 \lambda^2 b_3^2 - b_4^2 b_3^2 + 6\lambda^2 b_2^2 + 16\lambda^2 - 8 b_3^2}{b_3 b_4 b_5 b_6} \right)\mathcal{O}_2 \notag \\
    &+ \left(\frac{2\sqrt{2}(6\lambda^2 b_2^2 + 80 \lambda^2 - 24 \lambda^4 -8 b_3^2 - 10 \lambda^2 b_3^2)}{b_2 b_3 b_4 b_5 b_6} \right)\mathcal{O}_0 -\lambda^3 \left(\frac{24 b_3^2 - 256}{b_2 b_3 b_4 b_5 b_6} \right)\left( \sigma^{y}_1 \sigma^{y}_2 - \sigma^{x}_1 \sigma^{x}_2 \right) \notag \\
    &+ \left(\frac{384\lambda^2 - 28\lambda^2 b_3^2 - 32\lambda^4}{b_2 b_3 b_4 b_5 b_6} \right)\left(\sigma^{x}_0 \sigma^{z}_1 \sigma^{x}_2 + \sigma^{x}_1 \sigma^{z}_2 \sigma^{x}_3 \right) + \frac{48 \lambda^5}{b_2 b_3 b_4 b_5 b_6}\left(\sigma^{x}_0 \sigma^{x}_1 + \sigma^{x}_2 \sigma^{x}_3 \right) \notag \\
    &+ \left(\frac{512\lambda - 32 \lambda b_3^2 +32\lambda^3}{b_2 b_3 b_4 b_5 b_6}\right)\sigma^{y}_1 \sigma^{y}_2 - \left(\frac{512\lambda - 32\lambda b_3^2 + 32\lambda^3}{b_2 b_3 b_4 b_5 b_6} \right)\sigma^{x}_0 \sigma^{z}_1 \sigma^{z}_2 \sigma^{x}_3
\end{align}
\end{widetext}

The Lanczos coefficients obtained corresponding to these Krylov vectors are given as follows
\begin{align}
    b_{1} &= 2 \sqrt {2} \\
    b_2 &= \sqrt{8 + 10 \lambda^2} \\
    b_3 &= \sqrt{\frac{2 \lambda ^2 \left(9 \lambda ^2+32\right)}{5 \lambda ^2+4}} \\
    b_4 &= 2 \sqrt{\frac{72 \lambda ^6+202 \lambda ^4+448 \lambda ^2+512}{45 \lambda ^4+196 \lambda ^2+128}} \\
    b_5 &= 2 \lambda  \sqrt{\frac{(5 \lambda ^2+4 )(16 - 9 \lambda^2)^2}{\left(9 \lambda ^2+32\right) \left(36 \lambda ^6+101 \lambda ^4+224 \lambda ^2+256\right)}} \\
    b_6 &= 4 \lambda  \sqrt{\frac{\left(9 \lambda ^2+32\right) \left(\lambda ^4+2 \lambda ^2+4\right)}{36 \lambda ^6+101 \lambda ^4+224 \lambda ^2+256}}
\end{align}
Using \eqref{mateqn} now allows us to obtain the expression \eqref{AGP4b}.

\section{Gauge constraint for regularized AGP}
\label{appC}

In this section, we shall derive the gauge condition satisfied by the regularized AGP operator. The regularised AGP operator may be written as
\begin{align}
	A_\lambda = -\frac{1}{2}\int_{-\infty}^{\infty}\sg{t}e^{-\mu |t|}\partial_\lambda H (t) \mathrm{d}t
\end{align}
Calculating the expectation value between two eigenstates $\bra{m}$ and $\ket{n}$ (with $m \neq n$) gives us the expression
\begin{align}
	\bra{m}A_\lambda \ket{n} = -\frac{i \omega_{m n}}{\mu^2 + \omega_{m n}^{2}}\bra{m}\partial_\lambda H \ket{n}
\end{align}
where $\omega_{m n} = E_{m} - E_{n}$.
We note the following
\begin{align}
	\omega_{m n}^2 \bra{m}A_\lambda \ket{n} = \bra{m} [H,[H,A_\lambda]] \ket{n} \notag \\
	-i\omega_{m n}\bra{m}\partial_\lambda H \ket{n} = -i\bra{m}[H,\partial_\lambda H]\ket{n}
\end{align}
Therefore, using it, one can note that this expression reduces to
\begin{align}
	\bra{m} [H, i\partial_\lambda H + [H, A_\lambda]] + \mu^2 A_\lambda \ket{n} = 0
\end{align}
In matrix notation, this expression can be written as
\begin{align}
	[H, [H, i\partial_\lambda H + [H, A_\lambda]] + \mu^2 A_\lambda ] = 0
\end{align}
However, this gauge constraint is redundant in the sense that it generates an over-complete set of equations for the coefficients $\alpha$. The reason behind it is the fact that the following equation holds
\begin{align}
	[H, i\partial_\lambda H + [H, A_\lambda]] + \mu^2 A_\lambda = 0
\end{align}
which can be seen by using \eqref{AGPC4} and \eqref{arec}.

\section{Integrable non-interacting model}
\label{appD}
We evaluate an approximate analytical expression for the AGP for the following system
\begin{align}
	H = \frac{1}{2}\sum_{i}\sigma^{x}_{i}\sigma^{x}_{i + 1}  + h \sigma^{z}_{i}\label{IsingChainAtCrit}
\end{align}
This is an Ising chain. We shall evaluate the AGP at criticality (i.e., $h = 1$). The explicit AGP operator has been derived by a Jordan-Wigner transformation to a free fermion chain \cite{PhysRevLett.109.115703}. It has also been derived via the Lanczos approach \cite{Takahashi:2023nkt}. In this section, we use the autocorrelation function to derive the AGP norm. We use the following result \cite{brandt1976exact} for the Ising chain at criticality.
\begin{align}
	&\mathcal{C}_{l, m}(t) \equiv (\sigma^{z}_{l}(t) \sigma^{z}_{m})\notag\\ &= (J_{2 l - 2 m}(2 t))^{2} - J_{2 l - 2 m + 1}(2 t)J_{2 l - 2 m - 1}(2 t)\label{autoCIsing}
\end{align}
where $J_{\nu}(t)$ is the Bessel function of first kind.
The autocorrelation function for the deformation operator $\sum_{i}\sigma^{z}_{i}$ is
\begin{align}
	\mathcal{C}(t) = \sum_{l = 1}^{L}\sum_{m = 1}^{L}\mathcal{C}_{l, m}(t)
\end{align}
To evaluate the AGP norm, we need the following integrals
\begin{align}
	&\mathbf{I}_{1}(\nu) = \int_{0}^{\infty}J_{\nu}(2 t)^{2}e^{-\mu t}\mathrm{d}t \notag\\
	&= \frac{16^{\nu } \mu ^{-2 \nu -1} \Gamma \left(\nu +\frac{1}{2}\right)^2 \, _2\tilde{F}_1\left(\nu +\frac{1}{2},\nu +\frac{1}{2};2 \nu +1;-\frac{16}{\mu ^2}\right)}{\pi } 
\end{align}
and
\begin{align}
	&\mathbf{I}_{2}(\nu) = \int_{0}^{\infty}J_{\nu-1}(2 t)J_{\nu + 1}(2 t)e^{-\mu t}\mathrm{d}t 
	= ((2 \nu )!)^2 \mu ^{-2 \nu -1}\notag\\ &\times \, _4\tilde{F}_3\left(\nu +\frac{1}{2},\nu +\frac{1}{2},\nu +1,\nu +1;\nu ,\nu +2,2 \nu +1;-\frac{16}{\mu ^2}\right)
\end{align}
where $\, _p\tilde{F}_q$ is the regularized hypergeometric function with $p,q$ parameters \cite{weisstein_2003}.

The regulated AGP norm is written as
\begin{align}
	\vert \vert A_{\lambda}(1) \vert \vert^2 = \sum_{l, m} \left(\frac{1}{\mu} + \partial_\mu\right)(\mathbf{I}_{1}(2 l - 2 m) - \mathbf{I}_{2}(2 l - 2 m))
\end{align}
which evaluates to (where $\nu = 2 l - 2 m$)
\begin{align}
	&\vert \vert A_\lambda(1) \vert \vert^2 =\sum_{l, m = 1}^{L}\frac{(\nu +1) (2 \nu )! \mu ^{-2 (\nu +2)}}{\Gamma (\nu +2)^2}\notag\\
	&\times \Bigg\{-\mu ^2 \nu  \, _3F_2\left(\nu +\frac{1}{2},\nu +\frac{1}{2},\nu +1;\nu +2,2 \nu +1;-\frac{16}{\mu ^2}\right) \notag\\
	&- \frac{4 (\nu +1) (2 \nu +1)}{\nu +2}\notag\\ &\times\, _3F_2\left(\nu +\frac{3}{2},\nu +\frac{3}{2},\nu +2;\nu +3,2 \nu +2;-\frac{16}{\mu ^2}\right) \Bigg\}\label{AGPIsingAnalytic}
\end{align}
This sum is highly non-trivial, so we leave the expression as it is. At least formally, this presents an analytical expression for the AGP norm of the Ising chain \eqref{IsingChainAtCrit} at criticality $h = 1$. However, it is important to note that this expression is approximate. That is due to the fact (as explained in detail in \cite{brandt1976exact}) that the autocorrelation function \eqref{autoCIsing} holds only for $l, m$ away from the boundary of the chain. Since that is not the case here, some of the terms in the sum \eqref{AGPIsingAnalytic} will pick up errors.

\bibliography{references.bib}
\end{document}